\documentclass{iopjournal}
\usepackage{amsmath}
\newtheorem{theorem}{Theorem}
\usepackage{amssymb}
\usepackage{bm}
\usepackage{rotating}
\usepackage{tabularray}
\usepackage{amsmath, amssymb, bm}
\usepackage{cite}

\begin{document}

\articletype{Paper} %	 e.g. Paper, Letter, Topical Review...

\title{Optimal Estimation of Temperature in Finite-sized System}

\author{Shaoyong Zhang$^1$\orcid{0000-0000-0000-0000}, Zhaoyu Fei$^{1,*}$\orcid{0000-0000-0000-0000} and Xiaoguang Wang$^1$\orcid{0000-0000-0000-0000}}

\affil{$^1$Zhejiang Key Laboratory of Quantum State Control and Optical Field Manipulation, Department of Physics, Zhejiang Sci-Tech University, 310018 Hangzhou, China}

\affil{$^*$Author to whom any correspondence should be addressed.}

\email{feizhaoyu@zstu.edu.cn}

\keywords{temperature fluctuations, optimal estimation, energy-temperature uncertainty}

\begin{abstract}
Temperature of a finite-sized system fluctuates due to the thermal fluctuations. However, a systematic mathematical framework for measuring or estimating the temperature is still underdeveloped. Here, we incorporate the estimation theory in statistical inference to estimate the temperature of a finite-sized system and propose optimal estimation based on the uniform minimum variance unbiased estimation. Treating the finite-sized system as a thermometer measuring the temperature of a heat reservoir, we demonstrate that different optimal estimation of parameters yield different formulas of entropy, e.g., optimal estimation of inverse temperature (or temperature) aligns with the Boltzmann entropy (or Gibbs entropy). The optimal estimation leads to a achievable energy-temperature uncertainty relation and exhibits sample-size dependence, coinciding with their counterparts in nanothermodynamics. The achievable bound and the non-Gaussian distribution of temperature enable experimental testing in finite-sized systems.
\end{abstract}

\section{Introduction}

Temperature fluctuations become increasingly apparent as the system size decreases. Due to random heat transfer between a finite-sized system and its reservoir, their temperatures are not always equal. This behavior challenges the zeroth law of thermodynamics and the notion of a well-defined temperature for the finite-sized system. Experimentally, Gaussian-type temperature fluctuations have been observed in paramagnetic salt high resolution thermometer (HRT) \cite{bib39,day1997fluctuation,chui2001squid}, ferromagnetic alloy HRT \cite{bib40}, and nanoscale electron calorimeter \cite{bib41}. It is verified that the temperature of the system exhibits fluctuations, while the temperature of the reservoir remains fixed. 

Such challenges manifest in various research areas. (a) Definition of temperature in finite-sized systems including the definition of entropy and temperature in isolated systems \cite{bib4,bib6,bib8,campisi2005mechanical,campisi2025statistical}, the existence of negative absolute temperatures \cite{bib1,bib3,bib9,bib10,bib14,bib17}, the energy–temperature uncertainty relation (ETU) \cite{bohr1932faraday,bib52} and the thermodynamics of finite-sized systems \cite{bib33,bib35,bib36,bib37,bib38}; (b) Uncertainty of temperature measurement in quantum thermometry \cite{bib48,bib45,bib29,bib42,refree1.11,refree1.22} and nanoelectromechanical resonators \cite{refree1.1,refree1.2}; (c) Temperature fluctuations in nonequilibrium systems including superstatistics \cite{refree2.1,refree2.2,refree2.21}, nonequilibrium processes and stochastic thermodynamics \cite{salazar2016exactly,salazar2019stochastic,chen2022microscopic,fei2024temperature,finite}. Despite considerable progress, conclusions drawn from specific viewpoints or scenarios are sometimes inconsistent, and a unified understanding of thermal fluctuations in finite-sized systems remains incomplete. This situation reflects the absence of a general mathematical framework for studying such fluctuations, which motivates our work to develop a more systematic approach.

Historically, Einstein firstly investigated equilibrium fluctuations \cite{mishin2015thermodynamic,einstein1907limit,einstein1910theorie,landau2013statistical} by relating the probability of macroscopic states to entropy through an inversion of the Boltzmann's entropy formula \footnote{The operation of inverting formula of entropy has been further developed and extended to Tsallis statistics to describe nonequilibrium fluctuations of strongly correlated systems ~\cite{refree1.3}.}. Bohr suggested that there should exist a form of complementarity between temperature and energy in thermodynamics similar to that of position and momentum in quantum theory \cite{bohr1932faraday,bib52}. Then, Mandelbrot linked the ETU ($\beta$ denotes the inverse temperature of the system and $E$ denotes the energy of the system)
\begin{equation} 
	\Delta \beta \Delta E   \geq  1
\end{equation}
to the Cramér–Rao bound in estimation theory through the polemic exchange with Feshbach and Kittel \cite{feshbach1987small,kittel1988temperature,mandelbrot1989temperature,kittel1989gibbs}, where $\Delta(\cdot)=\sqrt{\langle \cdot^{2}\rangle -\langle \cdot\rangle ^{2}} $. This practice originated from 1956 \cite{mandelbrot1956} when Mandelbrot applied the estimation theory to the derivation of statistical thermodynamics from purely phenomenological principles (also see \cite{bib25,bib26,falcioni2011estimate}). It inspires us to use the uncertainty of temperature measurement to understand temperature fluctuations in finite-sized systems. For more dicussions about thermodynamic uncertainty relations, we refer to \cite{uffink1999thermodynamic}.

In this article, aiming to extend the study beyond the Gaussian-type fluctuation near the thermodynamic limit, we develop a general mathematical framework for thermal fluctuations based on estimation theory. By proposing the concept, optimal estimation of temperature, we rigorously
describe temperature fluctuations in a finite-sized system within the framework of statistical inference. This approach provides a systematic perspective for studying the thermodynamics of finite-sized systems.

In estimation theory, unbiasedness and efficiency are the two important criteria for evaluating estimators.
An estimator $\hat{\theta}$ of parameter $\theta$ is called “unbiased", if its mean value satisfies \cite{casella2024statistical}
\begin{equation} \label{unbiased}
	\langle  \hat{\theta}\rangle =\theta ,\quad\text{for}\quad \theta \in \Theta,
\end{equation}
where $\Theta$ is the range of all possible parameter values. If the mean squared errors of two estimators $\hat{\theta}_{1}$ and $\hat{\theta}_{2}$ satisfy
\begin{equation} \label{efficient}
	\langle (\hat{\theta}_{1}-\theta)^{2}\rangle    \leq \langle (\hat{\theta}_{2}-\theta)^{2}\rangle   ,   \quad \text{for}\quad \theta \in \Theta,
\end{equation}
then $\hat{\theta}_{1}$ is called to be more efficient than $\hat{\theta}_{2}$ \cite{casella2024statistical}.
In experimental terms, unbiasedness corresponds to accuracy (the absence of systematic error), while efficiency corresponds to precision (the minimization of statistical uncertainty). Therefore, an optimal estimator is the most efficient estimator among all unbiased estimators for any $\theta \in \Theta$, i.e., the uniform minimum variance unbiased estimation (UMVUE).

For systems without upper energy bound, we find a one-to-one link between Boltzmann (Gibbs) entropy $S_{\text{B}}=\ln [\sigma(E)\epsilon  ]$ $(S_{\text{G}}=\ln \Omega(E))$ and the optimal estimation of inverse temperature (temperature), i.e., Boltzmann temperature $\hat{\beta}_{B}\equiv\partial S_{B}/\partial E$ (Gibbs temperature $\hat{T}_{G}\equiv(\partial S_{G}/\partial E)^{-1}$). Here, $\sigma(E)$ is the density of the states of the system at energy $E$, $\epsilon$ is the width of a narrow energy interval, and $\Omega(E)=\int_{E_{\text{min}}}^{E}\text{d}E^{'}\sigma(E^{'})$ is the volume in phase space enclosed by a hypersurface of constant energy $E$ \cite{bib54}, where $E_{\text{min}}$ denotes the lower bound of system's energy and we set Boltzmann constant $k_{B}=1$. Hence, different formulas of entropy apply to optimal estimation of different parameters, rather than being contradictory. For systems with upper energy bound, the Boltzmann temperature still yields optimal estimation regardless of whether the temperature is positive or negative, while the Gibbs temperature yields optimal estimation only for positive temperatures.

The variance of the optimal estimator of temperature is the infimum among the variances of all unbiased estimators of temperature, thus yielding an achievable ETU. In the large-$N$ limit, it leads to a refined ETU
\begin{equation} 
	\Delta \beta \Delta E   \geq\Delta \hat{\beta}_{B} \Delta E   =  1+\frac{1}{4}(\mu_{3}) ^{2},
\end{equation}
where $\mu_{3}=\langle  (E-\left\langle E\right\rangle)  ^{3}\rangle / (\Delta E)^{3}$ denotes the skewness of the canonical distribution, and $\hat{\beta}_{B}$ is shown to be the optimal estimator of the inverse temperature in the following paper.

In repeated sampling, the optimal estimation of temperature exhibits sample-size dependence, coinciding with their counterparts in nanothermodynamics \cite{bib33,bib35,bib36,bib37,bib38}. Actually, the “sampling ensemble” can be regarded as the statistical interpretation of the replica trick in Hill's nanothermodynamics. And a Gaussian-type temperature fluctuation naturally emerges from a general temperature distribution due to the central limit theorem (CLT).

Compared with quantum thermometry, which mainly focuses on improving the precision of temperature measurements (especially for low temperatures) through dynamical \cite{bib48,bib45,bib42,refree1.11} or steady-state properties\cite{bib29,refree1.1,refree1.11,refree1.2,refree1.22}, we aim to describe temperature fluctuations in finite-sized systems through a rigorous mathematical framework. Therefore, unlike thermometry, which typically seeks to improve precision by modifying system's energy spectrum, optimizing probes, or performing sufficiently many measurements, we focus on the optimal estimation of temperature for a given energy spectrum with a finite number of measurements. Moreover, the UMVUE derived in this work provides an achievable bound that is tighter than the Cramér–Rao bound, potentially shedding light on the generation of more accurate methods of temperature estimation.

\section{Estimation of temperature in single sampling} \label{section2}

Developing Mandelbrot's viewpoint about temperature fluctuations \cite{mandelbrot1989temperature,mandelbrot1956,bib25,bib26}, i.e., “temperature for systems-in-isolation should be viewed as a statistical estimate of the parameter of a conjectural canonical distribution", we treat a finite-sized system in canonical ensemble as a thermometer of a reservoir. The $N$-particle system contacts a real or hypothetical  heat reservoir with temperature $T$. At equilibrium, the probability of the system at state $i$ obeys the Boltzmann-Gibbs formula:
\begin{equation}\label{BGD}
	P_{i}=\frac{d_{i}}{Z}e^{-\beta E_{i}},
\end{equation}
where $\beta=T^{-1}$ denotes the inverse temperature, $E_{i}$ $(d_{i})$ denotes the energy (degeneracy) of energy level with label $i\in D$, $D$ denotes the set of energy level indices, and $Z=\sum_{i\in D}d_{i}e^{-\beta E_{i}}$ is the partition function. 

Let $E_{\text{max}}$ denote the upper bound of system's energy, $\hat{\beta}^{(k)}$ ($\hat{T}^{(k)}$) denote the unbiased estimator of $\beta^{k}$ ($T^{k}$), where $k$ is an integer. We discuss the unbiased estimation of temperature in two cases. 

Firstly, when $E_{\text{max}}=\infty$, one option of $\hat{\beta}^{(k)} $ reads (see Appendix \ref{betak}):
\begin{equation} \label{option}
	\hat{\beta}^{(k)}=
	\begin{cases}
		\frac{\text{d}^{k}_{E_{i}}\sigma(E_{i})}{\sigma(E_{i})} & \text{if } \sigma(E_{i})\neq0 \\
		0   & \text{if } \sigma(E_{i})=0
	\end{cases},
\end{equation}
where $\sigma(E)
\equiv\sum_{i\in D}\delta(E-E_{i})d_{i}$ denotes the density of states and \footnote{Eq.~\eqref{dek} was noted by Gibbs \cite{bib56}, but its connection to estimation theory was not explored.}
\begin{equation} \label{dek}
	\text{d}^{k}_{E}\sigma(E)\equiv
	\begin{cases}
		\frac{\text{d}^{k}\sigma(E)}{\text{d}E^{k}} & \text{for } k \textgreater 0 \\
		\sigma(E) & \text{for } k = 0 \\
		\int_{E_{\text{min}}}^{E_{}}\text{d}E^{'}\text{d}_{E'}^{k+1}\sigma(E')   & \text{for } k\textless 0
	\end{cases}
	.
\end{equation}
For $k=1$, we write $\text{d}^{1}_{E}\sigma(E)$ as $\text{d}_{E}\sigma(E)$  for simplicity. In the derivation, we have assumed the density of states $\sigma(E)$ satisfies
\begin{equation} \label{assumption}
	\left. \text{d}^{l}_{E}\sigma(E)\right| _{E=E_{\text{min}}}=0
\end{equation}
for integer $l\in[0,k) $, which holds in classical systems but not in quantum systems. This is relevant to the third law of thermodynamics \footnote{The third law of thermodynamics is a macroscopic manifestation of quantum effects \cite{bib62,fermi2012thermodynamics,sewell2002quantum}. One formulation of the third law is $\displaystyle\lim_{T \to 0} S=0$, 
	which corresponds to the system’s energy being $E_{\min}$. If $\sigma(E_{\min}) = 1$, the law holds; if $\sigma(E_{\min}) = 0$, the limit $\lim_{T \to 0} S$ does not exist, and the law fails.}.

\begin{figure}[t]
	\centering
	% 左图
	\begin{minipage}[b]{0.45\textwidth} % 左图宽度
		\centering
		\includegraphics[width=\textwidth]{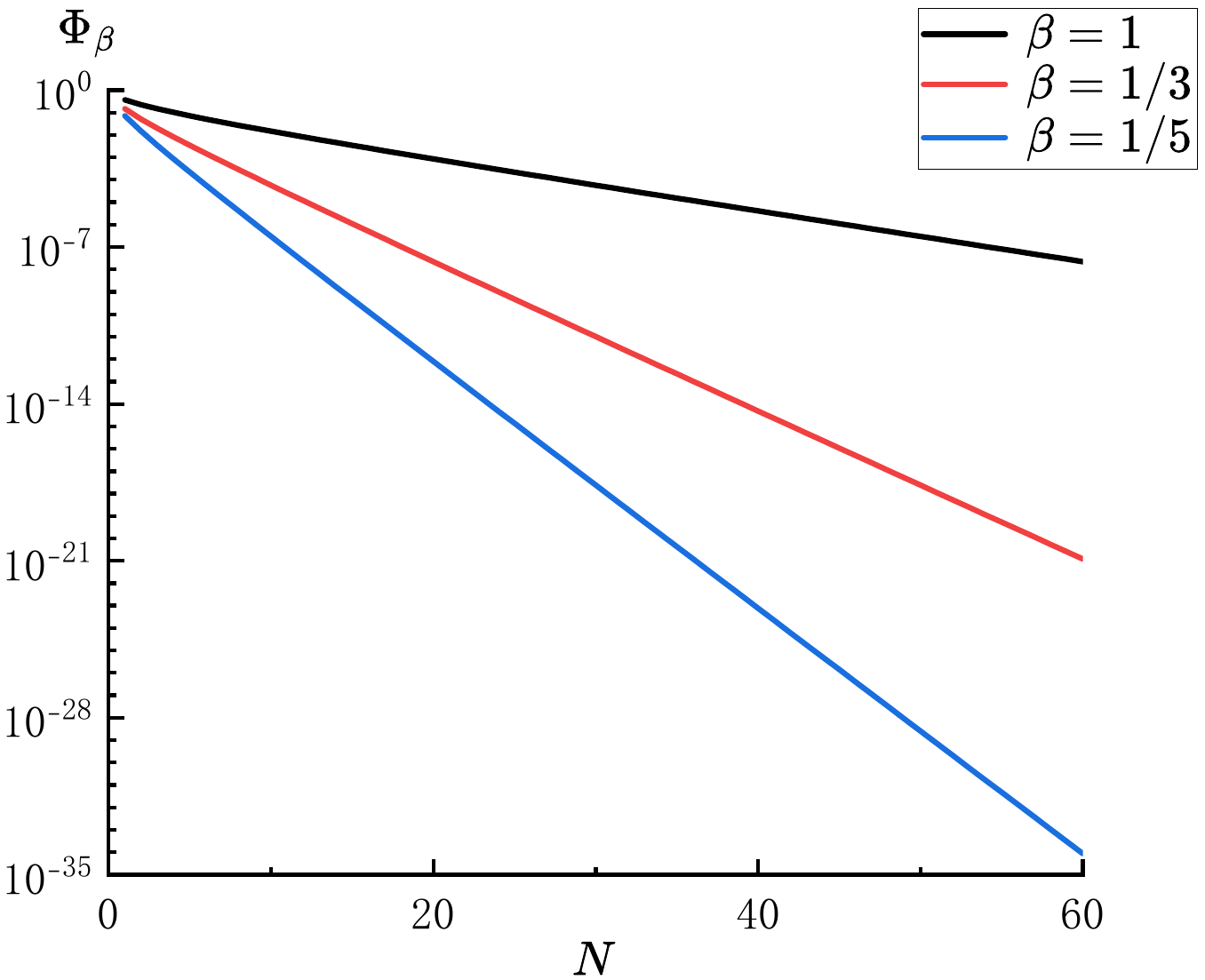}
		\caption{Relative bias of $\beta$ varies as $N$ when $E_{\text{max}}=\infty$. 
			Here, $\sigma(E)=e^{\sqrt{N(E-E_{\text{min}})}}$ is analogous to the density of states 
			of the low-temperature ideal Fermi gas (see Appendix \ref{fermi}), where the value of 
			$E_{\text{min}}$ does not affect the results.}
		\label{ER}
	\end{minipage}
	\hspace{0.1\textwidth} % <<< 调节这里控制右图向左移动
	% 右图
	\begin{minipage}[b]{0.42\textwidth} % 右图宽度
		\centering
		\includegraphics[width=\textwidth]{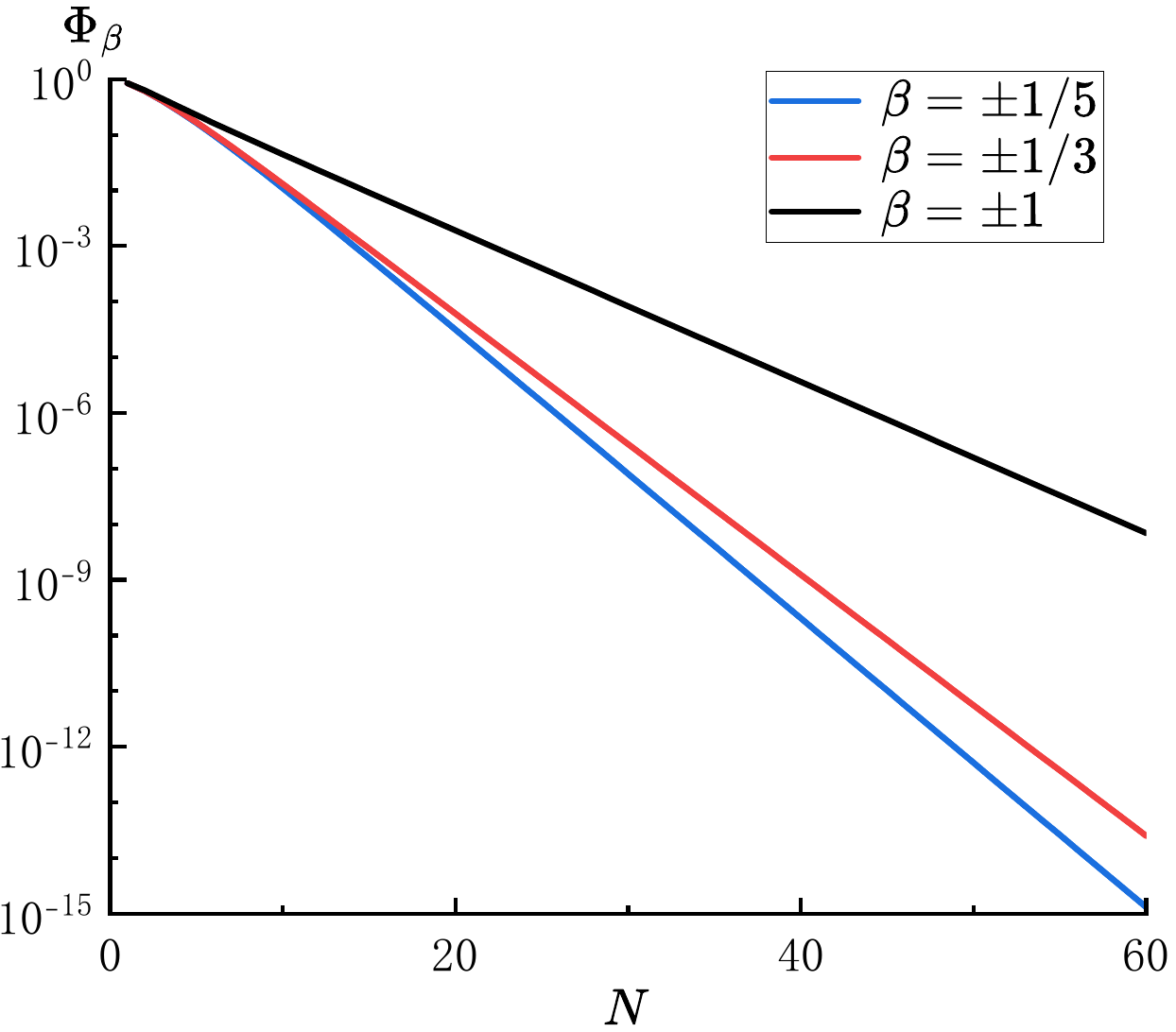}
		\caption{Relative bias of $\beta$ varies as $N$ when $E_{\text{max}}$ is finite. 
			Here, $\sigma(E)$ is the density of states of an $N$ non-interacting two-level system 
			(see Appendix \ref{twolevel}) and we set the spacing of energy levels $\varepsilon=1$.}
		\label{ERROR BETA1}
	\end{minipage}
\end{figure}

When the above assumption \eqref{assumption} is invalid, the estimator of $\beta^{k}$ for $k<0$, as well as that of $\beta$ given in Eq.~\eqref{option} are no longer unbiased and instead exhibit an exponentially decaying bias with $N$. In figure \ref{ER}, we illustrate this bias in an analogous low-temperature Fermi gas system using relative bias $\Phi_{\beta}=|\langle \hat{\beta}^{(1)}\rangle -\beta|/|\beta|$. Therefore, when $N$ is sufficiently large, the bias becomes negligible, and Eq.~\eqref{option} still provides their unbiased estimators.

It is worth noting that $k$ imposed by the above estimators have an upper bound (in this paper we mainly consider the case $k=\pm1$), depending on the degrees of freedom of the system. As a example, taking the density of states $\sigma(E)=E^{3N/2-1}$ of a classical ideal gas as an example \cite{bib62}, $\hat{\beta}^{(k)}$ in Eq. \eqref{option} is
\begin{equation}
	\hat{\beta}^{(k)}=\frac{\Gamma\left( \frac{3N}{2}\right) }{\Gamma\left( \frac{3N}{2}-k\right)} E^{-k},
\end{equation}
where $k\leq3N/2-2$ and $\Gamma$ denotes the Gamma function \footnote{When {$k \ge 3N/2 - 1$}, the proof in Appendix~\ref{betak} used to verify that {$\hat{\beta}^{(k)}$} is an unbiased estimator of {$\beta^{k}$} no longer holds.}.

When considering the density of states of a quantum many-body system, an issue arises of how to define the derivatives and integrals in Eqs. \eqref{option} and \eqref{dek} since the density of states is discrete. In practice, this is resolved by approximating the discrete density of states with a continuous function, as commonly done in statistical mechanics. For instance, in Section 1.4 of the textbook \cite{bib55}, the author carefully discusses how to approximate the density of states and the phase volume of a quantum ideal gas using a continuous function, as well as the effect of boundary conditions on the approximation.

Secondly, when $E_{\text{max}}$ is finite, the estimators of the powers of inverse temperature $(k>0)$ given in Eq.~\eqref{option} are no longer unbiased and instead exhibit a bias that decays exponentially with $N$. In figure \ref{ERROR BETA1}, we illustrate this bias in an $N$ non-interacting two-level systems, showing that it leads to a relative bias $\Phi _{\beta}\sim \text{e}^{-\alpha_{1} N}$, where the decay rate $\alpha_{1}= \ln(1+\text{e}^{-|\beta|\varepsilon})$ (see Appendix \ref{twolevel}), and $\varepsilon$ denotes the spacing of energy levels. Therefore, when $N$ is sufficiently large, the bias becomes negligible, and Eq.~\eqref{option} still provides their unbiased estimators.

\begin{figure}[t]
	\centering
	% 左图，左移 0.02\textwidth
	\hspace{-0.08\textwidth} 
	\begin{minipage}[b]{0.48\textwidth}
		\centering
		\includegraphics[width=\textwidth]{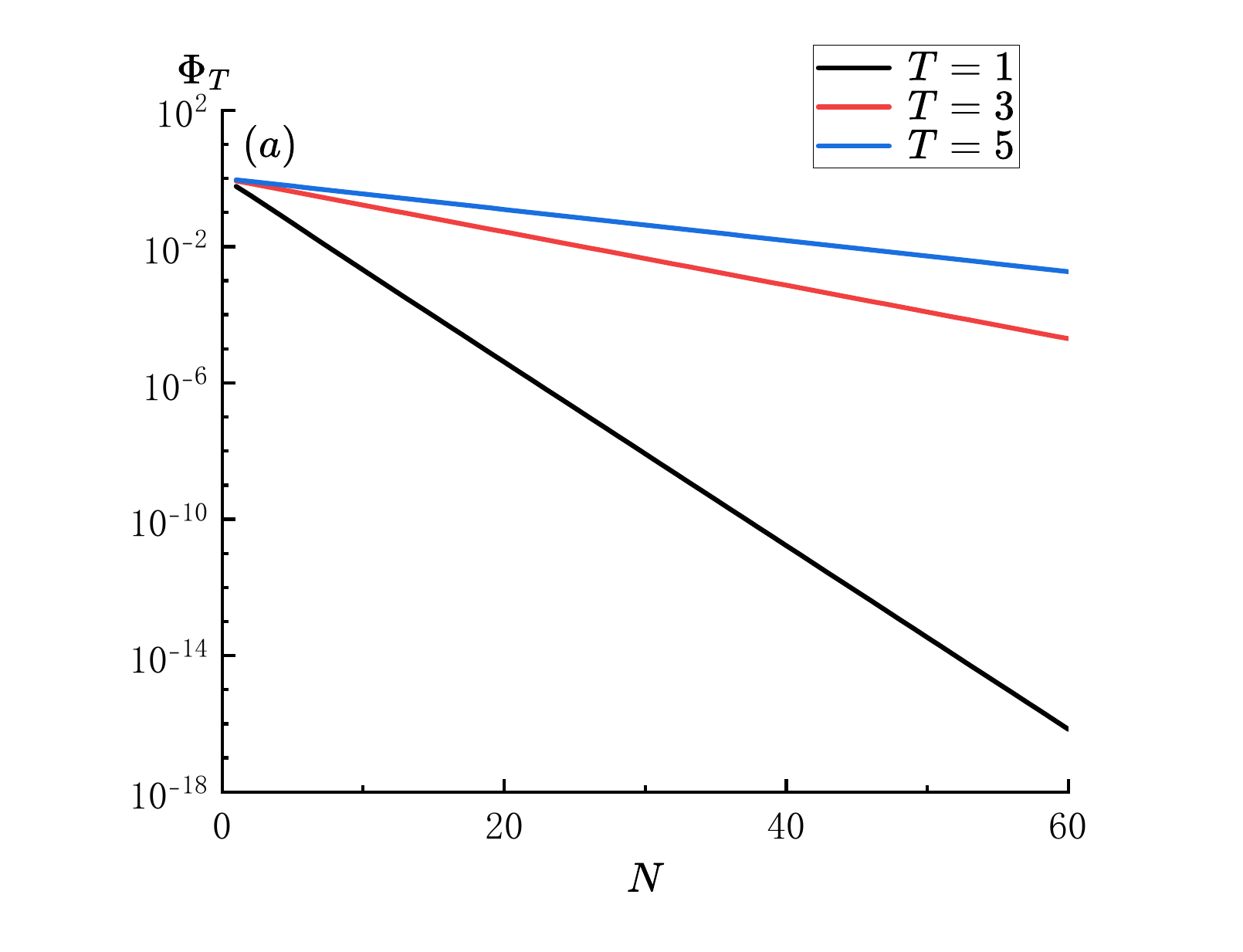}
	\end{minipage}
	\hspace{0.0\textwidth} % 右图间距
	% 右图
	\begin{minipage}[b]{0.48\textwidth}
		\centering
		\includegraphics[width=\textwidth]{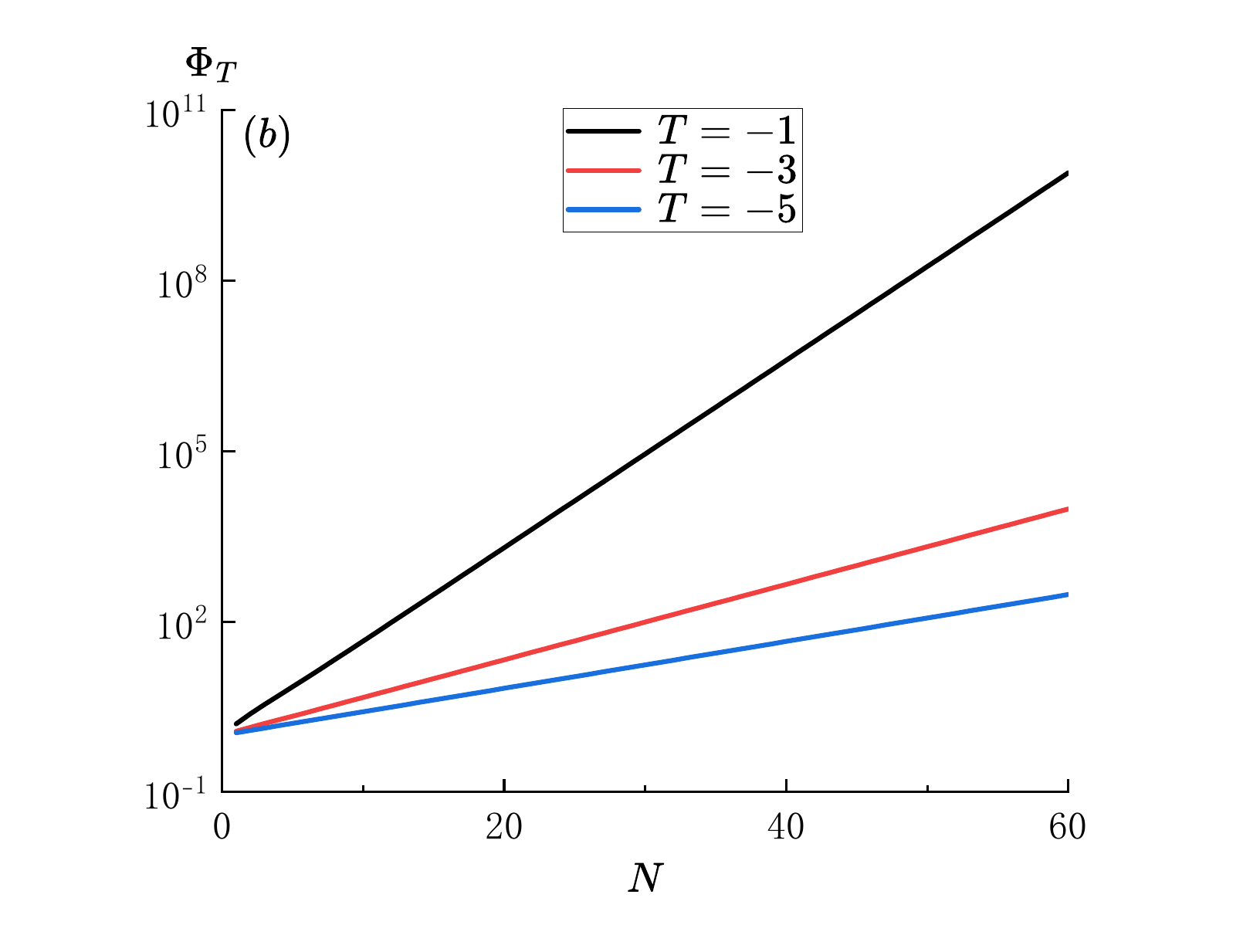}
	\end{minipage}
	
	\caption{Relative bias of $T$ varies as $N$ when $E_{\text{max}}$ is finite. 
		Here, the density of states is the same as figure~\ref{ERROR BETA1} (see Appendix \ref{twolevel}). 
		(a) $T>0$. (b) $T<0$.}
	\label{ERROR T}
\end{figure}

For the estimators of the powers of temperature $(k<0)$, we have to consider the cases separately. For positive temperatures, the estimators given by Eq. \eqref{option} exhibit a analogous bias that decays exponentially with $N$, which becomes negligible once 
$N$ is sufficiently large. In contrast, for negative temperatures, using Eq. \eqref{option} to estimate the system’s temperature results in a bias that grows exponentially with $N$, which demonstrates that Gibbs temperature cannot be used to estimate the temperature of a negative-temperature system. In figure \ref{ERROR T}, we illustrate these biases in an $N$ non-interacting two-level systems, showing that they lead to a relative bias $ \Phi_{T}=|\langle \hat{\beta}^{(-1)}\rangle -T|/|T|\sim \text{e}^{-\alpha_{2} N}$, where the decay (increase) rate  $\alpha_{2}=\frac{\beta \varepsilon}{2}+\ln \cosh (\frac{\beta\varepsilon}{2})$ (see Appendix \ref{twolevel}). The above indicates that an optimal estimator for the temperature of a negative-temperature system likely does not exist. Moreover, we extend the unbiased estimator of $\beta^{\alpha}$ to $\alpha\in \mathbb{R}$ (see Appendix \ref{betaalpha}).

In the following, we prove one of the main results of this paper: the unbiased estimators in Eq.~\eqref{option} are optimal, i.e., the UMVUEs of temperature in canonical ensemble. First, we introduce an essential theorem in statistical inference, the \text{Rao-Blackwell}-\text{Lehmann}-\text{Scheff\'{e} theorem} (RBLS) \cite{casella2024statistical,kay1993fundamentals,lehmann2006theory}:
\begin{theorem}
	If $\hat{\theta}$ is an unbiased estimator of $\theta$ and $T$ is a sufficient statistic for $\theta$, The expectation of $\hat{\theta}$ conditional on $T$, i.e., $\tilde{\theta}= \langle\hat{\theta} |  T\rangle  $ is
	
	\item1. a vaild estimator for $\theta$ (not dependent on $\theta$)\\
	2. unbiased\\
	3. of lesser or equal variance than that of $\hat{\theta}$, for all $\theta$\\
	Additionally, if the sufficient statistic is complete, then the unbiased estimator is the UMVUE.
\end{theorem}
A corollary following the RBLS theorem \cite{casella2024statistical,kay1993fundamentals,lehmann2006theory} is: If $T$ is a complete sufficient statistic for the parameter $\theta$, then an estimator $\tilde{\theta}(T)$ that depends only on $T$ is the UMVUE of its expected value.

	 The Boltzmann–Gibbs formula in Eq.~\eqref{BGD} belongs to the exponential family in statistical inference, and its complete sufficient statistic is $E_{i}$ \cite{casella2024statistical}. According to the above corollary, since the estimators $\hat{\beta}^{(k)}$ in Eq.~\eqref{option}
are not only unbiased but also a functions of $E_{i}$, they are the UMVUEs of $\beta^{k}$. Moreover, no simple functional relationship exists among the UMVUEs of different moments of $\beta$ implies that knowing the optimal estimator of the temperature does not directly yield the estimator of any function of the temperature.

Especially, for $k=\pm1$, we have
\begin{equation} \label{bolz}
	\begin{split}
		\langle \hat{\beta}^{(1)}\rangle&=\langle \hat{\beta}_{B}\rangle=\beta. \\
		\langle \hat{\beta}^{(-1)}\rangle &=\langle  \hat{T}_{G}\rangle  =T.
	\end{split}
\end{equation}
Hence, the Boltzmann entropy $S_{B}$ and the Gibbs entropy $S_{G}$ correspond to the UMVUE of inverse temperature and temperature respectively. It is worth mentioning that $\hat{\beta}_{B}$ and $\hat{T}_{G}$
have been discussed in previous studies.  However, there are various controversies among these studies, mainly concerning the existence of negative temperatures \cite{bib1,bib3,bib9,bib10,bib14,bib17}, the validity of the equipartition theorem \cite{bib1,campisi2005mechanical,campisi2025statistical}, the requirements of the zeroth law of thermodynamics \cite{bib6,bib9}, the additivity of entropy, adiabatic invariance, and more \cite{bib6,bib8,campisi2005mechanical}. These controversies arise because most studies start from the four laws of thermodynamics and empirical facts valid in the thermodynamic limit. In contrast, we base our approach on the mathematical theorems in statistical inference, in which different estimators correspond to different formulas of entropy and temperature.

In our framework, the optimized scheme for obtaining the optimal estimation of temperature is as follows:
\begin{enumerate}
	\item 	
	
	Determine the expressions for $\sigma$ and $\Omega$, which are either known before the measurement or can be obtained using the method in Section \ref{section large}. This method relates $\sigma$ to the partition function and its derivatives in the large-$N$ limit, allowing us to calculate $\sigma$ and $\Omega$ once the partition function is known.
	\item 
	Once the expressions for $\sigma$ and $\Omega$ are available, the expression for the estimator of temperature (Eqs. \eqref{option} and \eqref{dek}) can be determined.
	\item 
	Measure the energy of the system and substitute it into Eqs. \eqref{option} and \eqref{dek} to obtain the optimal estimation of the temperature.
\end{enumerate} 
The optimal estimate of temperature represents the temperature of the finite-sized system in this measurement, following Mandelbrot's viewpoint \cite{mandelbrot1989temperature,mandelbrot1956,bib25,bib26}. In quantum thermometry, it has been demonstrated that energy measurements are the most informative measurement~\cite{refree1.11}. It is worth noting that temperature fluctuations may also arise from reservoirs with finite capacity (see Ref.~\cite{finite} for details).

\section{Estimation of temperature for large-$N$ systems} \label{section large}
The UMVUE of $T$ and $\beta$, $\hat{T}_{G}$ and $\hat{\beta}_{B}$, depend on the expression of $\sigma(E)$. For large-$N$ systems, we approximate the expression of $\sigma(E)$ through the partition function $Z$ and its derivatives, which can be obtained through measuring heat capacity of the system. Using the Laplace approximation (saddle point approximation) \cite{bib60,bib61} or the Darwin–Fowler method in Sec. 9.1 of \cite{bib62} (see Appendix \ref{largen}), we obtain
\begin{equation} \label{entropy12}
	\ln \sigma(\left\langle E\right\rangle )=S-\ln \sqrt{ 2\pi\left( -\frac{\partial\left\langle E\right\rangle}{\partial \beta}\right) } +O(N^{-1}),
\end{equation}
where $S=\ln Z-\beta  \partial \ln Z/ \partial \beta $ is the thermodynamics entropy in a canonical ensemble.

Taking the derivative of Eq. \eqref{entropy12} with respect to $E$ on both sides, we have
\begin{equation} \label{BL3}
	\hat{\beta}_{B}(E)=\hat{\beta}_{L}(E)+\frac{1}{2}\frac{\text{d}^{2}_{E}\hat{\beta}_{L}(E)}{\text{d}_{E}\hat{\beta}_{L}(E)}+O(N^{-2}),
\end{equation}
where $ \hat{\beta}_{L}(E)$ (given by $-\left. \partial\ln Z/\partial\beta \right| _{\beta=\hat{\beta}_{L}(E)}=E $) is the the first-order moment (also the maximum likelihood) estimator of $\beta$.

Because $\hat{\beta}_{B}$ is the UMVUE of $\beta$, its variance is the infimum of among variances of unbiased estimators of $\beta$, i.e., 
\begin{equation} \label{Dbetayange}
	\Delta \hat{\beta} \Delta E   \geq \Delta \hat{\beta}_{B} \Delta E\geq1 ,
\end{equation}
where the variance of energy in the canonical ensemble $(\Delta E)^{2}=T^{2}C$ is also the Fisher information of parameter $\beta$ and $C$ denotes the heat capacity. Eq. \eqref{Dbetayange} is the achievable ETU for finite $N$, which is tighter than the usual ETU.

\begin{figure}[t]
	\centering
	
	\hspace{-0.02\textwidth}  % 设置两个图之间的空白
	
	\includegraphics[scale=0.3]{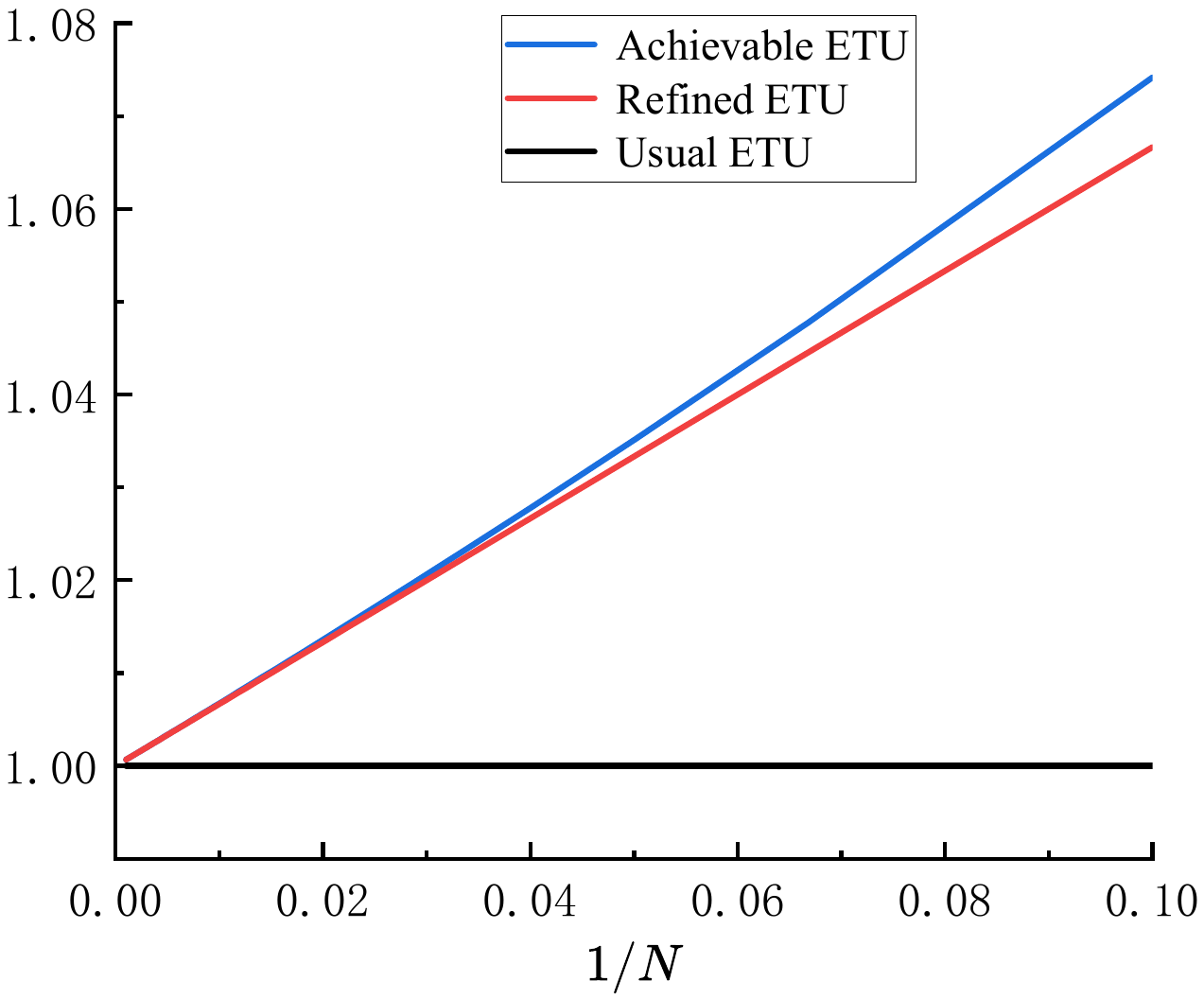}  % 调整宽度

	\caption{ETU varies as $1/N$. Here, $\sigma(E)=E^{\frac{3N}{2}-1}$ is the density of states of the classical ideal gas (see Appendix \ref{classical}). We set the inverse temperature $\beta=1$.}
	\label{DBETA}
\end{figure}

In the large-$N$ limit, the optimal estimation of temperature leads to a refined ETU using Eq.~\eqref{BL3} (see Appendix \ref{largen}):

\begin{equation} \label{Dbeta}
	\Delta \hat{\beta} \Delta E   \geq \Delta \hat{\beta}_{B} \Delta E  = 1+\frac{1}{4}(\mu_{3}) ^{2}+O(N^{-2}).
\end{equation}
In figure \ref{DBETA}, the achievable ETU approach the refined ETU as $N$ increases.  As $N\to \infty$, the skewness approaches 0 due to the CLT. The Gaussian-type temperature fluctuation and the usual ETU are recovered.\\

	Similarly, $\hat{T}_{G}$ and $\hat{T}_{L}=\hat{\beta}_{L}(E)^{-1}$ satisfy
\begin{equation} \label{GL3}
	\hat{T}_{G}(E)=\hat{T}_{L}(E)-\frac{\hat{T}_{L}(E)^{2}}{2}\frac{\text{d}^{2}_{E}\hat{T}_{L}(E)}{\text{d}_{E}\hat{T}_{L}(E)}+O(N^{-2}).
\end{equation}
And we have
\begin{equation}\label{DT33}
	\frac{\Delta \hat{T} \Delta E}{T^{2}}   \geq \frac{\Delta \hat{T}_{G} \Delta E}{T^{2}} \geq 1.
\end{equation}
The optimal estimation of temperature leads to a refined $\Delta T \Delta E/T^{2}$ in the large-$N$ limit (see Appendix \ref{largen}):
\begin{equation}\label{DT}
	\frac{\Delta \hat{T} \Delta E}{T^{2}}   \geq \frac{\Delta \hat{T}_{G} \Delta E}{T^{2}}  = 1+\frac{T^{2}}{4C^{3}}\left( \frac{\text{d}C}{\text{d}T}\right) ^{2}+O(N^{-2}).
\end{equation}
As an illustration, the estimation of temperature in classical ideal gas is considered (see Appendix \ref{classical}).

\section{Estimation of temperature in repeated sampling}
In single sampling, the measured energy of the finite-sized system exhibits significant fluctuations. To improve the precision of measurement, we discuss the UMVUE of temperature in repeated sampling with the sample size $M$. Let $\bm{i}$=$\left( i_{1},i_{2},...,i_{M}\right) $ denote the microstates in repeated sampling. The $M$-independent sampling is equivalent to a single sampling of $M$-identical non-interacting systems. Here, we call the $M$-identical non-interacting systems “$M$-sampling ensemble”.

At equilibrium, the probability of the system at state $\bm{i}$ still obeys the Boltzmann-Gibbs formula:
\begin{equation}\label{PM}
	P_{\bm{i}}=\frac{d_{\bm{i}}}{Z^{M}} \text{e}^{-\beta E_{\bm{i}}} ,
\end{equation}
where $ E_{\bm{i}}=\sum_{n=1}^{M}E_{i_{n}}\quad(d_{\bm{i}}=\prod_{n=1}^{M} d_{i_{n}})$ denotes the total energy (degeneracy) of state $\bm{i}$.

Similar to the single-sampling case, the unbiased estimator of $\beta^k$ with sample size $M$ is

\begin{equation} \label{option M}        
	\hat{\beta}_{M}^{(k)}=
	\begin{cases}
		\frac{\text{d}^{k}_{E_{\bm{i}}}\sigma_{M}\left( E_{\bm{i}}\right)  }{\sigma_{M}\left( E_{\bm{i}} \right) } & \text{if } \sigma_{M}\left( E_{\bm{i}}\right) \neq0 \\
		0   & \text{if } \sigma_{M}\left( E_{\bm{i}}\right) =0
	\end{cases},
\end{equation}
where $\sigma_{M}(E)\equiv\sum_{\bm{i}}\delta\left( E- E_{\bm{i}}\right)d_{\bm{i}} $ denotes the density of states of the sampling ensemble. And Eq.~\eqref{option M} is the UMVUE of $\beta^{k}$ in repeated sampling for system without upper energy bound, according to the RBLS theorem \cite{casella2024statistical,kay1993fundamentals,lehmann2006theory}. For system with upper energy bound, the result is similar to that obtained earlier. Especially, for $k=\pm1$, we have
\begin{align}
	\langle \hat{\beta}_{M}^{(1)}\rangle&=\langle \hat{\beta}_{B,M}\rangle =\left\langle   \frac{\partial \ln (\sigma_{M}\epsilon)}{\partial E}\right\rangle  =\beta, \\
	\langle \hat{\beta}_{M}^{(-1)}\rangle &=\langle \hat{T}_{G,M}\rangle =\left\langle  \left(   \frac{\partial \ln \Omega_{M}}{\partial E}\right) ^{-1}\right\rangle    =T,
\end{align}
which motivates us to generalize the definitions of Boltzmann entropy and Gibbs entropy in the “$M$-sampling ensemble”,
\begin{align}
	S_{B,M}(E)&=\ln [ \sigma_{M}(E)\epsilon] , \\
	S_{G,M}(E)&=\ln \Omega_{M}(E),
\end{align}
where $\Omega_{M}(E)=\int_{E_{\text{min}}}^{E}\sigma_{M}(E')\text{d}E'$ is the volume in phase space enclosed by a hypersurface of constant energy $E$ in the $M$-sampling ensemble.

For the sampling ensemble, the $M$-dependence of density of states, entropy, and temperature implies that sample size should be treated as a independent state variable.
In the large-$M$ limit, the sampling ensemble corresponds to Hill's nanothermodynamics with $M$-copies, as realized through the replica trick \cite{bib33,bib35,bib36,bib37,bib38}. Accordingly, thermodynamic quantities of entropy and temperature coincide with their counterparts in nanothermodynamics.

Because $\hat{\beta}_{B,M}$ is the UMVUE of $\beta$ in $M$ samplings, its variance is the infimum of among variances of unbiased estimators of $\beta$, i.e., 
\begin{equation} \label{Machie}
	\sqrt{M}\Delta \hat{\beta}_{M} \Delta E   \geq \sqrt{M}\Delta \hat{\beta}_{B,M} \Delta E\geq1 ,
\end{equation}
where $\Delta E=\Delta E_{\bm{i}}/\sqrt{M}$ denotes the standard deviation of energy in the single sampling. Eq. \eqref{Machie} gives the achievable ETU in $M$ samplings. For large but finite $M$, the optimal estimation of temperature leads to a refined ETU following the similar procedure in Appendix \ref{largen}, 
\begin{equation}
	\begin{split}
		\sqrt{M}\Delta \hat{\beta}_{M} \Delta E   &\geq\sqrt{M}\Delta \hat{\beta}_{B,M} \Delta E = 1 + \frac{1}{4M}  (\mu_{3})^{2} + O\left( (MN)^{-2} \right).
	\end{split}	
\end{equation}
Similarly, we obtain 
\begin{equation}
	\frac{\sqrt{M}\Delta \hat{T}_{M} \Delta E}{T^{2}} \geq \frac{\sqrt{M}\Delta \hat{T}_{G,M} \Delta E}{T^{2}} \geq 1,
\end{equation}
and 	
\begin{equation}
	\begin{split}
		\frac{\sqrt{M}\Delta \hat{T}_{M} \Delta E}{T^{2}} &\geq\frac{\sqrt{M}\Delta \hat{T}_{G,M} \Delta E}{T^{2}}= 1 +   \frac{T^{2}}{4MC^{3}}\left( \frac{\text{d}C}{\text{d}T}\right) ^{2} + O\left( (MN)^{-2} \right)
	\end{split}
\end{equation}
in the large $M$ limit. As $M\to \infty$, the Gaussian-type temperature fluctuation and the \text{Cram\'{e}r-Rao bound} are recovered due to the CLT. In Table \ref{tab1}, we also list the relevant quantities in the estimation of temperature for the single sampling and repeated sampling.

	\begin{sidewaystable*}
	\vspace{-3cm}
	\caption{Relevant quantities in estimation of temperature for the single sampling and the repeated sampling.}\label{tab1}
	\begin{tblr}{
			width = \textwidth,
			colspec = {Q[c,m,3.6cm] Q[l,m,9cm] Q[l,m,10.5cm]},
			rowhead =3,
			hlines,
			vlines,
			rowsep = 12pt,
		}
		& \begin{center}
			The single sampling
		\end{center} & \begin{center}
			The repeated sampling
		\end{center} \\
		Partition function &
		$\ln Z = \ln \sigma(E^{*}) - \beta E^{*} + \ln \sqrt{\left. 2\pi\left( -\frac{\partial\hat{\beta}_{B}}{\partial E}\right)^{-1} \right| _{E=E^{*}} } + O(N^{-1})$ &
		$M\ln Z = \ln \sigma_{M}(E_{\bm{i}}^{*}) - \beta E_{\bm{i}}^{*} + \ln \sqrt{\left. 2\pi\left( -\frac{\partial\hat{\beta}_{B,M}}{\partial E}\right)^{-1} \right| _{E_{t}=E_{t}^{*}} } + O\left( (MN)^{-1} \right)$ \\
		Average energy &
		$\left\langle E \right\rangle = E^{*} + \left. \frac{1}{2} \frac{\partial^{2} \hat{\beta}_{B}}{\partial E^{2}} \right| _{E=E^{*}} \left( \frac{\partial E^{*}}{\partial \beta} \right)^{2} + O(N^{-1})$ &
		$M\left\langle E \right\rangle = E_{\bm{i}}^{*} + \left. \frac{1}{2} \frac{\partial^{2} \hat{\beta}_{B}}{\partial E_{\bm{i}}^{2}} \right| _{E_{\bm{i}}=E_{\bm{i}}^{*}} \left( \frac{\partial E_{\bm{i}}^{*}}{\partial \beta} \right)^{2} + O\left( (MN)^{-1} \right)$ \\
		The most probable energy &
		$E^{*} = \left\langle E \right\rangle - T + \frac{T^{2}}{2} \frac{\partial \ln C}{\partial T} + O(N^{-1})$ &
		$E_{\bm{i}}^{*} = M \left\langle E \right\rangle - T + \frac{T^{2}}{2} \frac{\partial \ln C}{\partial T} + O\left( (MN)^{-1} \right)$ \\
		Relation between canonical ensemble entropy and microcanonical ensemble entropy &
		$\ln \sigma(\left\langle E \right\rangle) = S - \ln \sqrt{ 2\pi \left( - \frac{\partial \left\langle E \right\rangle}{\partial \beta} \right) } + O(N^{-1})$ &
		$\ln \sigma_{M}(M \left\langle E \right\rangle) = MS - \ln \sqrt{ 2\pi M \left( - \frac{\partial \left\langle E \right\rangle}{\partial \beta} \right) } + O\left( (MN)^{-1} \right)$ \\
		Relation between the UMVUE and the maximum likelihood estimator of the inverse temperature &
		$\hat{\beta}_{B}(E) = \hat{\beta}_{L}(E) + \frac{1}{2} \frac{d^{2}_{E} \hat{\beta}_{L}(E)}{d_{E} \hat{\beta}_{L}(E)} + O(N^{-2})$ &
		$\hat{\beta}_{B,M}(E) = \hat{\beta}_{L}(E) + \frac{1}{2M} \frac{d^{2}_{E} \hat{\beta}_{L}(E)}{d_{E} \hat{\beta}_{L}(E)} + O\left( (MN)^{-2} \right)$ \\
		Refined ETU &
		$\Delta \hat{\beta}_{B} \Delta E = 1 + \frac{1}{4} (\mu_{3})^{2} + O(N^{-2})$ &
		$\sqrt{M} \Delta \hat{\beta}_{B,M} \Delta E = 1 + \frac{1}{4M}  (\mu_{3})^{2} + O\left( (MN)^{-2} \right)$ \\
		Refined $\Delta T \Delta E/T^{2}$ &
		$\frac{\Delta \hat{T}_{G} \Delta E}{T^{2}} = 1 + \frac{T^{2}}{4C^{3}}(\frac{\text{d}C}{\text{d}T})^{2} + O(N^{-2})$ &
		$\sqrt{M}\frac{\Delta \hat{T}_{G,M} \Delta E}{T^{2}} = 1 + \frac{T^{2}}{4MC^{3}}(\frac{\text{d}C}{\text{d}T})^{2} + O\left( (MN)^{-2} \right)$ \\
	\end{tblr}
\end{sidewaystable*}

\begin{figure}[t]
	\centering
	% 左图
	\begin{minipage}[b]{0.48\textwidth}
		\centering
		\includegraphics[width=\textwidth]{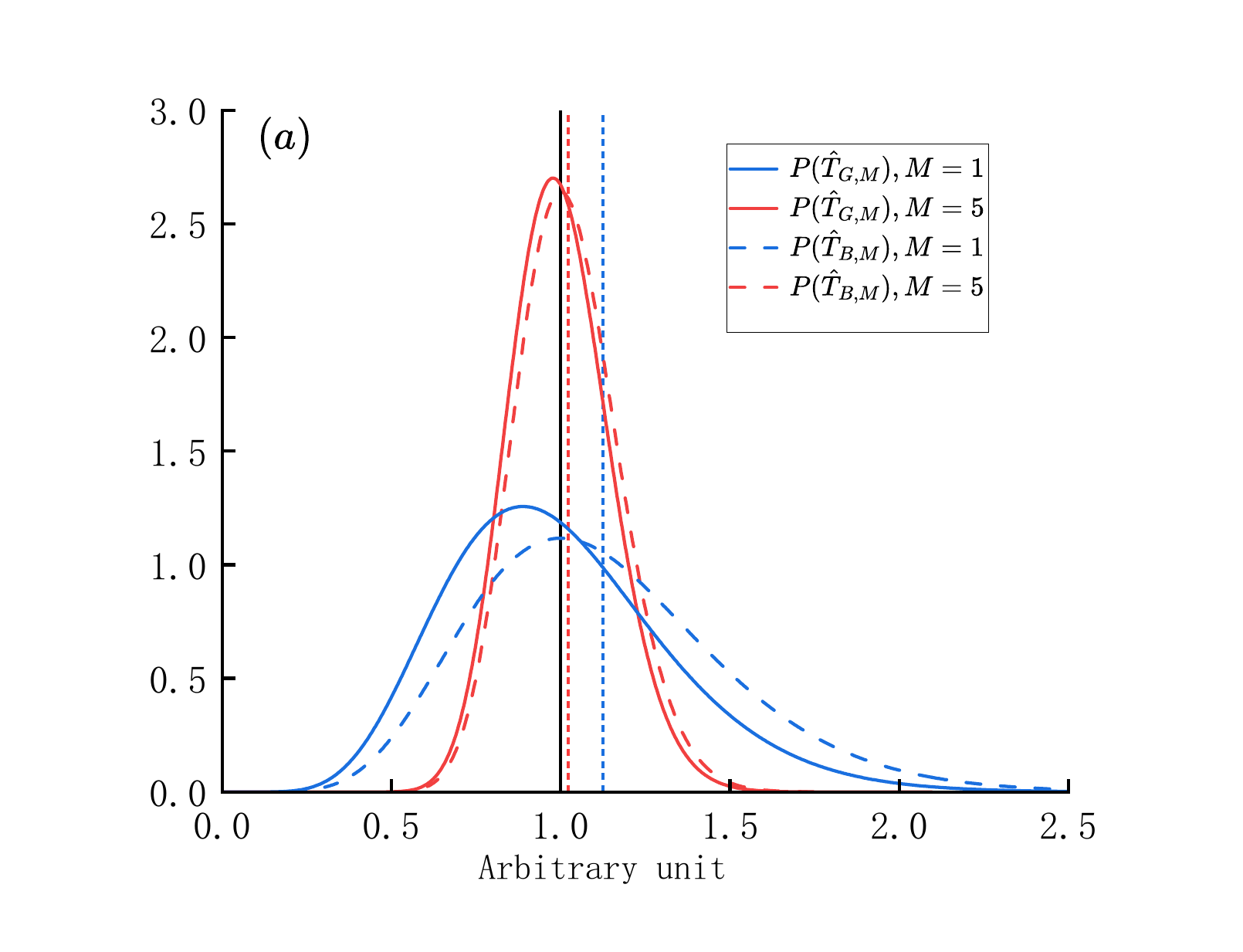}
		% 子标题，不占 figure numbering
	\end{minipage}
	\hfill
	% 右图
	\begin{minipage}[b]{0.48\textwidth}
		\centering
		\includegraphics[width=\textwidth]{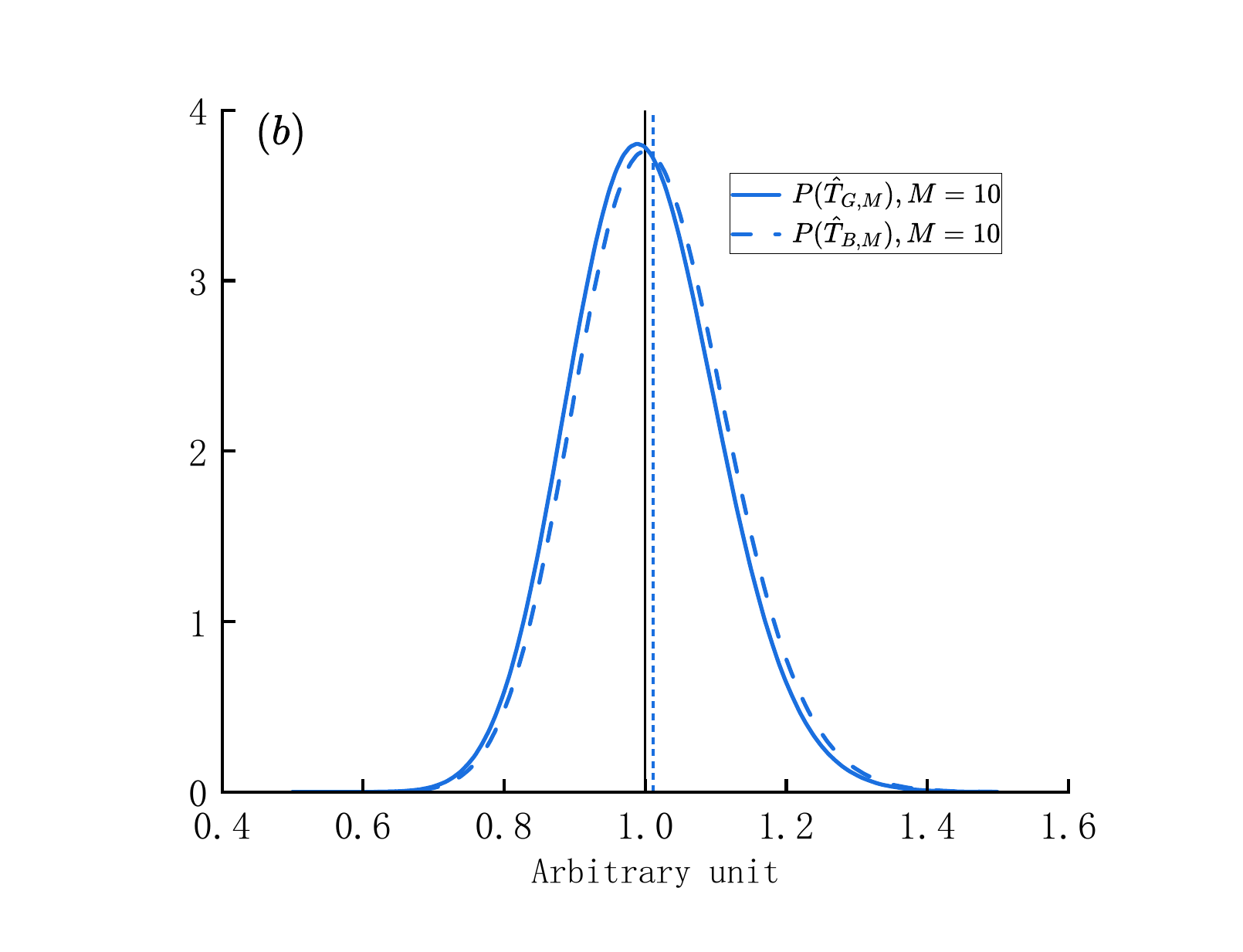}
		
	\end{minipage}
	
	\caption{Probability distributions $P$ of the estimators $\hat{T}_{G,M}$ and $\hat{T}_{B,M}$ for different $M$, where the vertical lines indicate the mean values of the corresponding distributions. 
		Here, $\sigma_{M}(E)=E^{\frac{3NM}{2}-1}$ is the density of states of the classical ideal gas \cite{bib62}. 
		We set temperature $T=1$ and particle number $N=6$.}
	\label{PT1015}
\end{figure}

In figure \ref{PT1015}, we show the probability distributions of the estimators $\hat{T}_{G,M}$ and $\hat{T}_{B,M}=(\partial \ln [\sigma_{M}(E)\epsilon]/\partial E)^{-1}$ in finite-sized system for different $M$. For small $M$, they both follow a non-Gaussian distribution. The non-Gaussian probability distribution of $\hat{T}_{G,M}$ exactly corresponds to the quasi-equilibrium state in Ref.~\cite{fei2024temperature} (also see \cite{mishin2015thermodynamic,einstein1910theorie,landau2013statistical}), which can be regarded as a realization of superstatistics \cite{refree2.1,refree2.2,refree2.21} in finite-sized systems. As $M$ increases, a Gaussian distribution naturally emerges. Meanwhile, regardless of the value of $M$, the optimal estimation of temperature is always unbiased ( $\left\langle \hat{T}_{G,M}\right\rangle =1$, while $\left\langle \hat{T}_{B,M}\right\rangle \neq 1$). And for $M=10$, the variance of $\hat{T}_{G,10}$ are minimum among the set of $\lbrace \hat{T}_{G,1}, \hat{T}_{G,5}, \hat{T}_{G,10}\rbrace $. Hence, our work provides a foundation for further experimental studies on the non-Gaussian distribution of temperature and the thermodynamic relation between entropy and temperature in finite-sized systems.

\section{Conclusion and Outlook}
In this work, we develop a systematic mathematical framework for measuring temperature fluctuations in finite-sized systems using the parameter estimation theory. Using this framework, we obtained the optimal estimation of temperature. It reveals the one-to-one link between different formulas of entropy and optimal estimation of different parameters. The UMVUE of temperature provides us with an achievable ETU, which is tighter than the usual one. In the large-$N$ limit, the optimal estimation of temperature leads to a refined ETU, which is related to the skewness of the canonical ensemble. Finally, as the sample size grows, the non-Gaussian distribution of temperature gradually approaches a Gaussian distribution due to the CLT, while the optimal estimation of temperature remains both unbiased and efficient—regardless of system size. Also, we provide a repeated sampling interpretation for the temperature and entropy in Hill’s nanothermodynamics.

The achievable bound for the ETU, the non-Gaussian distribution of temperature and the thermodynamic relation between entropy and temperature enable experimental testing in finite-sized systems, such as neutral atom arrays \cite{ebadi2021quantum,bluvstein2024logical} and biochemical oscillators \cite{cao2015free,lee2018thermodynamic}. For thermometry, the optimal estimation of temperature may improve the validity of the measurement and the sampling efficiency in many-body systems with a finite number of measurements..

%
% Each of the commands below will create an unnumbered section with the appropriate heading.
% Remove any sections that are not relevant for your article.
% All sections except suppdata will be removed if the [anonymous] option is used.
% See iopjournal-guidelines.pdf for more information.
%

\ack{This work was supported by the Science Challenge Project (No.~TZ2025017), the Quantum Science and Technology-National Science and Technology Major Project (Grant No.2024ZD 0301000) , the National Natural Science Foundation of China (Grant No.~12405046), and the Science Foundation of Zhejiang Sci-Tech University (Grants No.~23062088-Y and No.~23062181-Y).}

\data{All data that support the findings of this study are included within the article \cite{Zhang2026}.}
% For more information on IOP Publishing's research data policy see: https://publishingsupport.iopscience.iop.org/questions/research-data/

\app{}	
\appendix
\renewcommand{\thesection}{\Alph{section}}
\renewcommand{\theequation}{\thesection\arabic{equation}}
\setcounter{equation}{0}  % 附录开始前将公式计数器清零
\section{Unbiased estimator of $\beta^{k}$} \label{betak}
    
We prove that $\hat{\beta}^{(k)}$ is an unbiased estimator of $\beta^{k}$. Let us consider the case $k> 0$. For $k=1$, where $\text{d}_{E_{i}}\sigma(E_{i})=\text{d}\sigma(E_{i})/\text{d}E_{i}$, we have \cite{bib65}
\begin{equation}
	\begin{split}
		\langle \hat{\beta}^{(1)}\rangle
		&=\frac{1}{Z}\sum_{i\in D}\frac{1}{\sigma(E_{i})}\frac{\text{d}\sigma(E_{i})}{\text{d}E_{i}}d_{i}\text{e}^{-\beta E_{i}}\\
		&=\frac{1}{Z}\sum_{i\in D}\frac{1}{\sigma(E_{i})}\frac{\text{d}\sigma(E_{i})}{\text{d}E_{i}}d_{i}\text{e}^{-\beta E_{i}}\int_{-\infty}^{\infty} \text{d}E \delta(E-E_{i}) \\
		&=\frac{1}{Z}\int_{-\infty}^{\infty}	\text{d}E\text{e}^{-\beta E} \frac{1}{\sigma(E)}\frac{\text{d}\sigma(E)}{\text{d}E}\sum_{i\in D} \delta(E-E_{i})d_{i}\\
		&=\frac{1}{Z}\int_{E_{\text{min}}}^{\infty}{\text{e}^{-\beta E}}\text{d}\sigma(E)\\
		&=\left. \frac{\text{e}^{-\beta E}\sigma(E)}{Z} \right| ^{\infty}_{E_{\text{min}}}-
		\frac{1}{Z}\int_{E_{\text{min}}}^{\infty}\sigma(E)\text{d}\text{e}^{-\beta E}\\
		&=\frac{1}{Z}\int_{E_{\text{min}}}^{\infty}\sigma(E)\beta \text{e}^{-\beta E}\text{d}E\\
		&=\beta ,
	\end{split}
\end{equation}
where $\sigma(E)=\sum_{i}\delta(E-E_{i})d_{i}$ denotes the density of state, and we require that $\sigma(E_{\text{min}})=0$. In the derivation, the lower limit of the integral changes from negative infinity to $E_{\text{min}}$ due to $\sigma(E)=0$ for $E<0$. Eq.~\eqref{dek} was noted by Gibbs \cite{bib56}, but its connection to estimation theory was not explored.

Suppose that for $k=n$, $n$ is a positive integer, we have $\langle \hat{\beta}^{(n)}\rangle =\beta^{n}$. Then, for $k=n+1$, we have $\text{d}^{n+1}_{E_{i}}\sigma(E_{i})=\text{d}\left[  \text{d}^{n}_{E_{i}}\sigma(E_{i})\right]  /\text{d}E_{i}$ and
\begin{equation} \label{ABETA}
	\begin{split}
		\langle \hat{\beta}^{(n+1)}\rangle 
		&=\frac{1}{Z}\sum_{i\in D}\frac{1}{\sigma(E_{i})}\frac{\text{d}\left[  \text{d}^{(n)}_{E_{i}}\sigma(E_{i})\right] }{\text{d}E_{i}}d_{i}\text{e}^{-\beta E_{i}}\\
		&=\frac{1}{Z}\int_{E_{\text{min}}}^{\infty}\text{e}^{-\beta E}\text{d}\left[  \text{d}^{n}_{E}\sigma(E)\right] \\
		&=\left. \frac{\text{e}^{-\beta E} \text{d}^{n}_{E}\sigma(E)}{Z} \right| ^{\infty}_{E_{\text{min}}}-\frac{1}{Z}\int_{E_{\text{min}}}^{\infty}\left[  \text{d}^{n}_{E}\sigma(E)\right] \text{d} \text{e}^{-\beta E}\\
		&=\beta \frac{1}{Z}\int_{E_{\text{min}}}^{\infty}\left[ \text{d}^{n}_{E}\sigma(E)\right] \text{e}^{-\beta E} \text{d}E\\
		&=\beta^{n+1},
	\end{split}
\end{equation}
where we require that $\left. \text{d}^{l}_{E}\sigma(E)\right| _{E=E_{\text{min}}}=0$ for integer $l\in \left[ 0,n-1\right] $. Therefore, by using the method of mathematical induction, we prove that  $\hat{\beta}^{(k)}=\text{d}^{k}_{E_{i}}\sigma(E_{i})/\sigma(E_{i})$ is an unbiased estimate of $\beta^{k}$ when $k>0$.

Next, let us consider the case $k< 0$. For $k=-1$, where $\text{d}_{E_{i}}^{-1}\sigma(E_{i})=\int_{E_{\text{min}}}^{E_{i}}\text{d}E'\sigma(E')$, we have
\begin{equation} \label{AT}
	\begin{split}
		\langle \hat{\beta}^{(-1)}\rangle 
		=&\frac{1}{Z}\sum_{i\in D}\frac{\text{d}_{E_{i}}^{-1}\sigma(E_{i})}{\sigma(E_{i})}d_{i}\text{e}^{-\beta E_{i}} \\
		=&\frac{-1}{Z\beta}\int_{E_{\text{min}}}^{\infty}\left( \text{d}_{E}^{-1}\sigma(E)\right) \text{d}\text{e}^{-\beta E}\\
		=&\left. \frac{-\text{d}_{E}^{-1}\sigma(E)\text{e}^{-\beta E}}{Z\beta} \right| ^{\infty}_{E_{\text{min}}} +\frac{1}{Z}\int_{E_{\text{min}}}^{\infty}\frac{1}{\beta}\text{e}^{-\beta E}\text{d}\left( \text{d}_{E}^{-1}\sigma(E)\right)\\
		=&\frac{1}{\beta}\int_{E_{\text{min}}}^{\infty}\sigma(E)\frac{\text{e}^{-\beta E}}{Z}\text{d}E\\
		=&\beta^{-1}.
	\end{split}
\end{equation}
Follow the similar procedure, we prove that $\hat{\beta}^{(k)}=\text{d}_{E_{i}}^{k}\sigma(E_{i})/\sigma(E_{i})$ is an unbiased estimate of $\beta^{k}$ when $k<0$.\\

\section{The density of states of an analogous low-temperature ideal Fermi gas} \label{fermi}
\setcounter{equation}{0}
The internal energy of an low-temperature ideal Fermi gas is given by (see Section 8.1 of Ref.~\cite{bib55}) 

\begin{equation} \label{energy fermi}
	E=\frac{3}{5} N\varepsilon_{F} +\frac{\pi^{2}}{4}N\varepsilon_{F}\left( \frac{T}{\varepsilon_{F}}\right) ^{2}=E_{\text{min}}+aNT^{2},
\end{equation}
and the entropy is given by
\begin{equation} \label{entropy fermi}
	S=\frac{\pi^{2}TN}{2\varepsilon_{F}} =bNT,
\end{equation}
where $\varepsilon_{F}$ is generally referred to as the Fermi energy of the system, $a=\pi^{2}/4\varepsilon_{F}$ and $b=\pi^{2}/2\varepsilon_{F}$. Under the low-temperature approximation, $T$ is obtained from Eq.~\eqref{energy fermi}:

\begin{equation} \label{fermi t}
	T=\sqrt{\frac{E-E_{\text{min}}}{aN}}.
\end{equation}
Using the Boltzmann relation $S=\ln \sigma(E)$ and Eq.~\eqref{fermi t}, the density of states $\sigma(E)$ can be expressed as:
\begin{equation} \label{fermi sigma}
	\sigma(E)= \text{e}^{S}=\text{e}^{bNT}=\text{e}^{\frac{b}{\sqrt{a}}\sqrt{N(E-E_{\text{min}})}}.
\end{equation}
For simplicity, we set $b/\sqrt{a}=1$ when plotting figure~\ref{ER}. However, the qualitative results are unaffected regardless of the value of $b/\sqrt{a}$. The above approach provides an approximation for the density of states of a low-temperature ideal Fermi gas in the limit of $N\to\infty$. For finite $N$, a more detailed discussion of the density of states can be found in Refs.~\cite{bethe1936attempt,echter2025many}.

	\section{Relative biases of $\beta$ and $T$ in a two-level system} \label{twolevel}
	\setcounter{equation}{0}
	
	\subsection{Relative bias of $\beta$ in a two-level system}
\begin{figure}[t]
	\centering
	\begin{minipage}[b]{0.4\textwidth}
		\centering
		\includegraphics[width=\textwidth]{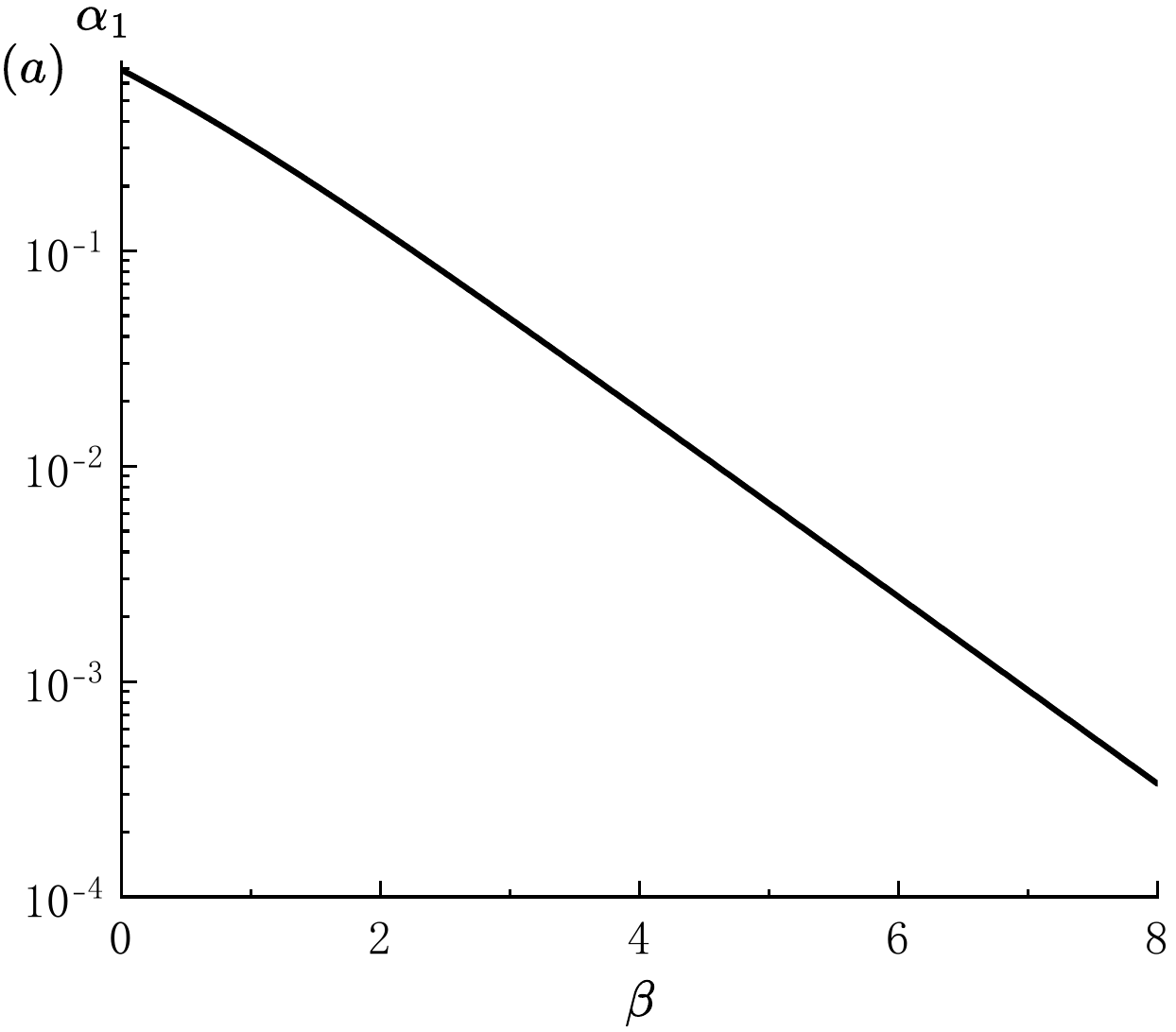}
	\end{minipage}
	\hspace{1.5cm} % 调整间距，比如0.3cm或1cm
	\begin{minipage}[b]{0.4\textwidth}
		\centering
		\includegraphics[width=\textwidth]{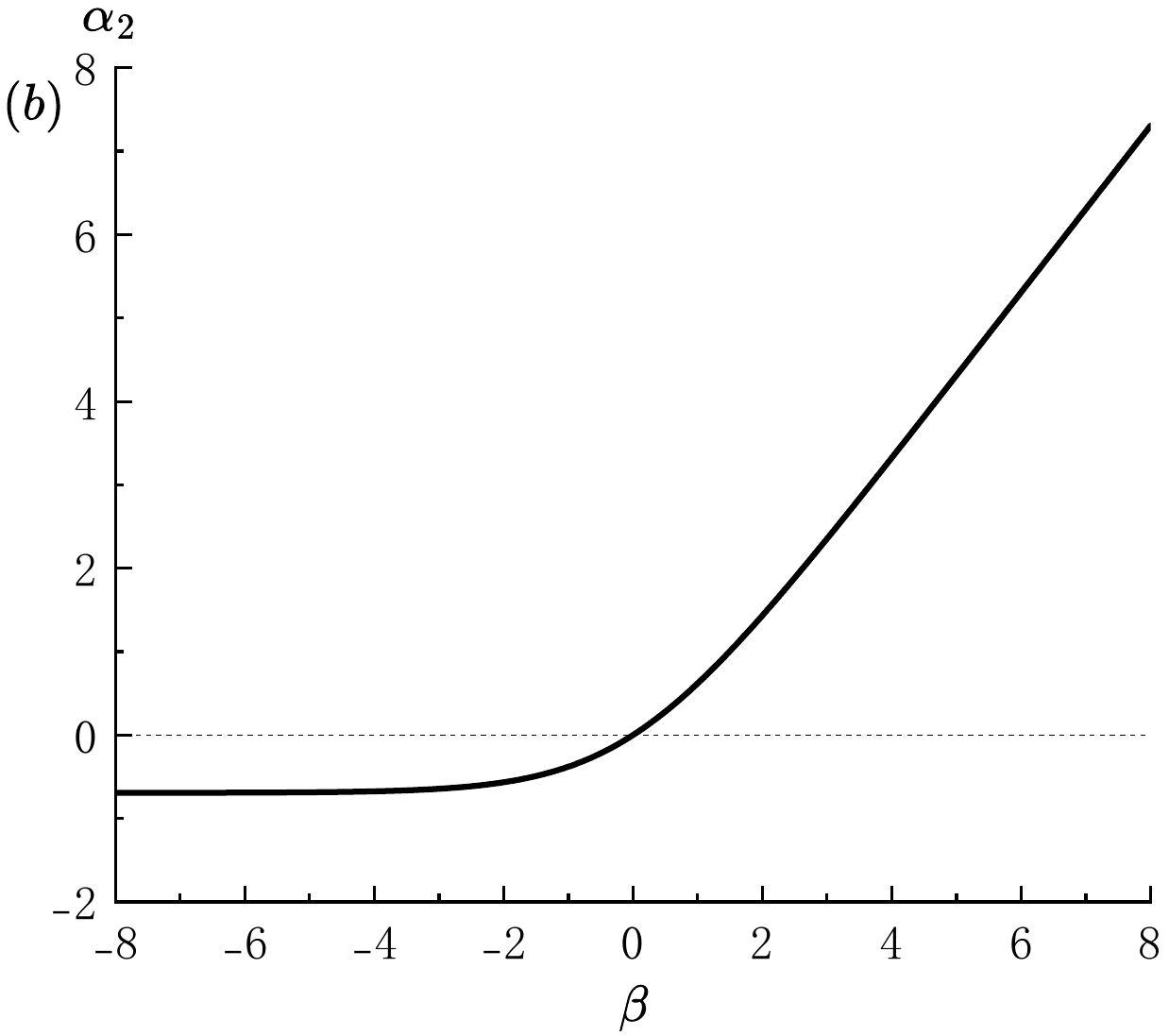}
	\end{minipage}
	
	\caption{Decay rate varies as $N$ when $E_{\text{max}}$ is finite. Here, the density of states of an $N$ non-interacting two-level system is Eq.~\eqref{density two-level} and we set the spacing of energy levels $\varepsilon=1$.} 
	\label{alpha}
\end{figure}

For systems with finite $E_{\text{max}}$, Eq.~(\ref{ABETA}) is modified to 
\begin{equation} \label{BBETA}
	\begin{split}
		\langle \hat{\beta}^{(1)}\rangle&=\left. \frac{\sigma(E)\text{e}^{-\beta E}}{Z} \right| ^{E_{\text{max}}}_{E_{\text{min}}}-
		\frac{1}{Z}\int_{E_{\text{min}}}^{E_{\text{max}}}\sigma(E)\text{d}\text{e}^{-\beta E}\\
		&=\frac{\sigma(E_{\text{max}})\text{e}^{-\beta E_{\text{max}}}-\sigma(E_{\text{min}})\text{e}^{-\beta E_{\text{min}}}}{Z}+\beta.
	\end{split}
\end{equation}
In an $N$ non-interacting two-level system, both the upper bound $E_{\text{max}}=N\varepsilon/2$ and lower bound $E_{\text{min}}=-N\varepsilon/2$ bounds of system's energy are finite. Here \cite{bib55},
\begin{equation} \label{density two-level}
	\sigma(E)=\frac{\Gamma(N+1)}{\varepsilon\Gamma\left( \frac{N}{2}-\frac{E}{\varepsilon}+1\right) \Gamma\left( \frac{N}{2}+\frac{E}{\varepsilon}+1\right) }
\end{equation}
and $Z=(\text{e}^{\beta\varepsilon/2}+\text{e}^{-\beta\varepsilon/2})^{N}$. Then, Eq.~\eqref{BBETA} is
\begin{equation} 
	\langle \hat{\beta}^{(1)}\rangle= \frac{\text{e}^{\frac{-\beta N \varepsilon}{2}}-\text{e}^{\frac{\beta N \varepsilon}{2}}}{\varepsilon\left( \text{e}^{\frac{\beta\varepsilon}{2}}+\text{e}^{\frac{-\beta\varepsilon}{2}}\right) ^{N}}+\beta ,
\end{equation}
And the relative bias of $\beta$ is

\begin{equation} 
	\begin{split}
		\Phi_{\beta}&=\left| \frac{\langle \hat{\beta}^{(1)}\rangle-\beta}{\beta}\right| \\
		&=\left| \frac{\text{e}^{\frac{-\beta N \varepsilon}{2}}-\text{e}^{\frac{\beta N \varepsilon}{2}}}{\beta\varepsilon\left( \text{e}^{\frac{\beta\varepsilon}{2}}+\text{e}^{\frac{-\beta\varepsilon}{2}}\right) ^{N}}\right| \\
		&\approx\frac{1}{\beta\varepsilon(1+\text{e}^{-|\beta|\varepsilon})^{N}}\\
		&=\frac{1}{\beta\varepsilon}\text{e}^{-N(\ln(1+\text{e}^{-|\beta|\varepsilon}))}\\
		&\sim \text{e}^{-\alpha_{1} N},
	\end{split}
\end{equation}
where the approximation used applies $\text{e}^{-N|\beta|\varepsilon}\ll 1$ with large $N$. In figure~\ref{alpha}a, $\alpha_{1}$ decreases exponentially with increasing inverse temperature. For a particular $N$, $\Phi_{\beta}$ is non-negligible, which indicates that the unbiased estimator $\hat{\beta}^{(1)}$ remains unbiased except at extremely low temperatures, as given in Eq.~\eqref{option}. This observation is consistent with previous results in quantum thermometry, where the estimation uncertainty diverges exponentially in the low-temperature regime for systems with finite energy gaps \cite{refree1.11}.

\subsection{Relative bias of $T$ in a two-level system}

For systems with finite $E_{\text{max}}$, Eq.~(\ref{AT}) is modified to
\begin{equation} \label{BT}
	\begin{split}
		\langle \hat{\beta}^{(-1)}\rangle 
		=&\left. \frac{-\text{d}_{E}^{-1}\sigma(E)\text{e}^{-\beta E}}{Z\beta} \right| ^{E_{\text{max}}}_{E_{\text{min}}} + \frac{1}{Z}\int_{E_{\text{min}}}^{E_{\text{max}}}\frac{1}{\beta}\text{e}^{-\beta E}\text{d}\left( \text{d}_{E}^{-1}\sigma(E)\right)\\
		=&\frac{\Omega(E_{\text{max}})\text{e}^{-\beta E_{\text{max}}}-\Omega(E_{\text{min}})\text{e}^{-\beta E_{\text{min}}}}{\beta Z}+\beta^{-1}.
	\end{split}
\end{equation}
In an $N$ non-interacting two-level system, we have $\Omega(E_{\text{min}})=\int_{E_{\text{min}}}^{E_{\text{min}}}\sigma(E)\text{d}E=0$ and $\Omega(E_{\text{max}})=\int_{E_{\text{min}}}^{E_{\text{max}}}\sigma(E)\text{d}E=Z|_{\beta=0}$. Then, with the help of Eq.~\eqref{density two-level}, Eq.~\eqref{BT} is
\begin{equation} 
	\langle \hat{\beta}^{(-1)}\rangle = \frac{2^{N}\text{e}^{\frac{-\beta N\varepsilon}{2}}}{\beta\left( \text{e}^{\frac{\beta\varepsilon}{2}}+\text{e}^{\frac{-\beta\varepsilon}{2}}\right) ^{N}}+\beta^{-1} ,	
\end{equation}
thus, the relative bias of $T$ is

\begin{equation} 
	\begin{split}
		\Phi_{T}&=\left| \frac{\langle \hat{\beta}^{(-1)}\rangle-T}{T}\right| \\
		&=\frac{2^{N}\text{e}^{\frac{-\beta N\varepsilon}{2}}}{(\text{e}^{\frac{\beta\varepsilon}{2}}+\text{e}^{\frac{-\beta\varepsilon}{2}})^{N}}\\
		&=\left( \frac{2\text{e}^{\frac{-\beta \varepsilon}{2}}}{\text{e}^{\frac{\beta\varepsilon}{2}}+\text{e}^{\frac{-\beta\varepsilon}{2}}}\right) ^{N}\\
		&= \frac{\text{e}^{\frac{-\beta N\varepsilon}{2}}}{\cosh (\frac{\beta \varepsilon}{2})^{N}}\\
		&=\text{e}^{-N(\frac{\beta\varepsilon}{2}+\ln \cosh(\frac{\beta\varepsilon}{2}))}\\
		&\sim \text{e}^{-\alpha_{2} N}.
	\end{split}
\end{equation}
In figure~\ref{alpha}b, when $T<0$, $\alpha_{2}$ becomes negative, causing $\Phi_{T}$ to increase exponentially, and the unbiased estimator of $\hat{\beta}^{(-1)}$ becomes invalid. When $T>0$, $\alpha_{2}$ is positive, and the unbiased estimator of $\hat{\beta}^{(-1)}$ is valid, except at extremely high temperatures where $\alpha_{2}$ approaches zero.

\section{Unbiased estimator of $\beta^{\alpha}$ for $\alpha\in\mathbb{R}$} \label{betaalpha}
\setcounter{equation}{0}
In this appendix, let us consider the unbiased estimation of the non-integer powers of temperature, i.e., $\langle \hat{\beta}^{(\alpha)}\rangle =\beta^{\alpha}$ for $\alpha \in R$. The expression of $\hat{\beta}^{(\alpha)}$ is still given by Eq.~\eqref{option} except that we extend the definition of $\text{d}^{\alpha}_{E}\sigma(E)$ in Eq.~\eqref{dek} in different cases. 

\subsection{Unbiased estimator of $\beta^{\alpha}$ for $\alpha<0$}

When $\alpha < 0$ , it is extended to
\begin{equation}\label{RLI}
	\text{d}^{\alpha}_{E}\sigma(E)=\frac{1}{\Gamma(-\alpha)}\int_{E_{\text{min}}}^{E}\sigma(u)\left( E-t\right) ^{-\alpha-1}\text{d}u,
\end{equation}
using the Riemann-Liouville integral \cite{bib58}. And we have
\begin{equation}
	\begin{split}
		\langle \hat{\beta}^{(\alpha)}\rangle
		&=\frac{1}{Z}\sum_{i\in D}\frac{1}{\sigma(E_{i})} \text{d}^{\alpha}_{E_{i}}\sigma(E_{i})d_{i}\text{e}^{-\beta E_{i}}\\
		&=\frac{1}{Z}\int_{E_{\text{min}}}^{\infty}\text{d}E \text{e}^{-\beta E} \text{d}_{E}^{\alpha} \sigma(E)\\
		&=\frac{1}{Z}\int_{E_{\text{min}}}^{\infty}\text{d}E \text{e}^{-\beta E} \frac{1}{\Gamma(-\alpha)}\int_{E_{\text{min}}}^{E}\text{d}t\sigma(t)\left( E-t\right) ^{-\alpha-1}\\
		&=\frac{1}{Z}\frac{1}{\Gamma(-\alpha)}\int_{E_{\text{min}}}^{\infty}\text{d}t \sigma(t) \int_{t}^{\infty}\text{d}E \text{e}^{-\beta E} \left( E-t\right) ^{-\alpha-1}\\
		&=\frac{1}{Z}\frac{\beta^{\alpha}}{\Gamma(-\alpha)}\int_{E_{\text{min}}}^{\infty}\text{d}t \sigma(t)\text{e}^{-\beta t}\int_{E_{\text{min}}}^{\infty}\text{d}x \text{e}^{-x}x^{-\alpha-1}\\
		&=\frac{1}{Z}\beta^{\alpha}\int_{E_{\text{min}}}^{\infty}\text{d}t \sigma(t)\text{e}^{-\beta t}\\
		&=\beta^{\alpha}.
	\end{split}
\end{equation}

\subsection{Unbiased estimator of $\beta^{\alpha}$ for $\alpha > 0$ and $n-1<\alpha<n$}

When $\alpha > 0$ and $n-1<\alpha<n$ for some integer $n$, it is extended to  
\begin{equation}\label{CFD}
	\text{d}^{\alpha}_{E}\sigma(E)=\frac{1}{\Gamma(n-\alpha)}\int_{E_{\text{min}}}^{E}\left( E-t\right) ^{n-\alpha-1}\text{d}^{n}_{u}\sigma(u)\text{d}u,	
\end{equation}
using the Caputo fractional derivative \cite{bib59} under the condition $\left. \text{d}_{E}^{l}\sigma(E)\right| _{E=0}=0$ for integer $l\in\left[  0,n-1\right] $. And we have
\begin{equation}
	\begin{split}
		\langle \hat{\beta}^{(\alpha)}\rangle
		=&\frac{1}{Z}\sum_{i\in D}\frac{1}{\sigma(E_{i})} \text{d}^{\alpha}_{E_{i}}\sigma(E_{i})d_{i}\text{e}^{-\beta E_{i}}\\
		=&\frac{1}{Z}\int_{E_{\text{min}}}^{\infty}\text{d}E \text{e}^{-\beta E} \text{d}_{E}^{\alpha} \sigma(E)\\
		=&\frac{1}{Z}\int_{E_{\text{min}}}^{\infty}\text{d}E \text{e}^{-\beta E} \frac{1}{\Gamma(n-\alpha)} \int_{E_{\text{min}}}^{E}\text{d}t\text{d}_{t}^{n}\sigma(t)\left( E-t\right) ^{n-\alpha-1}\\
		=&\frac{1}{Z}\frac{1}{\Gamma(n-\alpha)}\int_{E_{\text{min}}}^{\infty}\text{d}t \text{d}_{t}^{n}\sigma(t) \int_{t}^{\infty}\text{d}E \text{e}^{-\beta E} \left( E-t\right) ^{n-\alpha-1}\\
		=&\frac{1}{Z}\frac{\beta^{\alpha-n}}{\Gamma(n-\alpha)}\int_{E_{\text{min}}}^{\infty}\text{d}t\text{d}_{t}^{n}\sigma(t) \text{e}^{-\beta t}\int_{E_{\text{min}}}^{\infty}\text{d}x \text{e}^{-x}x^{-\alpha-1}\\
		=&\frac{1}{Z}\beta^{\alpha-n}\int_{E_{\text{min}}}^{\infty}\text{d}t \text{d}_{t}^{n}\sigma(t)\text{e}^{-\beta t}\\
		=&\beta^{\alpha}.
	\end{split}
\end{equation}
The last equlity is obtained by using  Eq. \eqref{ABETA}.

\subsection{Unbiased estimator of $\beta^{\alpha}$ for $\alpha\in\left[   0,1\right)$}

When $\alpha>0$ and the condition $\left. \text{d}_{E}^{l}\sigma(E)\right| _{E=0}=0$ is not satisfied for every $l\in\left[  0,n-1\right] $, we only consider the case $\alpha\in\left[   0,1\right)$. Here, it is extended to
\begin{equation} \label{RL}
	\text{d}^{\alpha}_{E}\sigma(E)=\frac{1}{\Gamma(1-\alpha)}\frac{\text{d}}{\text{d}E}\int_{E_{\text{min}}}^{E}\left( E-t\right) ^{-\alpha}\sigma(u)\text{d}u,	
\end{equation}
using the Riemann-Liouville fractional derivative \cite{bib58}. And we have
\begin{equation}
	\begin{split}
		\langle \hat{\beta}^{(\alpha)}\rangle
		&=\frac{1}{Z}\sum_{i\in D}\frac{1}{\sigma(E_{i})} \text{d}^{\alpha}_{E_{i}}\sigma(E_{i})d_{i}\text{e}^{-\beta E_{i}}\\
		&=\frac{1}{Z}\int_{E_{\text{min}}}^{\infty}\text{d}E \text{e}^{-\beta E} \text{d}_{E}^{\alpha} \sigma(E)\\
		&=\frac{1}{Z}\int_{E_{\text{min}}}^{\infty}\text{d}E \text{e}^{-\beta E} \frac{1}{\Gamma(1-\alpha)} \frac{\text{d}}{\text{d}E}\int_{E_{\text{min}}}^{E}\text{d}t\sigma(t)\left( E-t\right) ^{-\alpha}\\
		&=\frac{1}{Z}\left. \frac{1}{\Gamma(1-\alpha)}\text{e}^{-\beta E}\int_{E_{\text{min}}}^{E}\text{d}t\sigma(t)\left( E-t\right) ^{-\alpha}\right| ^{\infty}_{0}-\frac{1}{\Gamma(1-\alpha)}\int_{E_{\text{min}}}^{\infty}\text{d}\text{e}^{-\beta E}\int_{E_{\text{min}}}^{E}\text{d}t\sigma(t)\left( E-t\right) ^{-\alpha}\\
		&=\frac{1}{Z}\frac{\beta}{\Gamma(1-\alpha)}\int_{E_{\text{min}}}^{\infty}\text{d}t \sigma(t)\int_{t}^{\infty}\text{d}E \text{e}^{-\beta E} \left( E-t\right) ^{-\alpha}\\
		&=\frac{1}{Z}\frac{\beta^{\alpha}}{\Gamma(1-\alpha)}\int_{E_{\text{min}}}^{\infty}\text{d}t \sigma(t)\text{e}^{-\beta t}\int_{E_{\text{min}}}^{\infty}\text{d}x \text{e}^{-x}x^{-\alpha}\\
		&=\frac{1}{Z}\beta^{\alpha}\int_{E_{\text{min}}}^{\infty}\text{d}t \sigma(t)\text{e}^{-\beta t}\\
		&=\beta^{\alpha}.
	\end{split}
\end{equation}
We would like to emphasize that in the limit $\alpha \to 1^-$, $\lim_{\alpha \to 1^-}\langle \hat{\beta}^{(\alpha)}\rangle=\beta$, whereas $\langle\hat{\beta}^{(1)}\rangle=\beta+\sigma(E_{\text{min}})/Z$. This means the interchange of the limit $\alpha \to 1^-$ and the integral $\left\langle\cdot\right\rangle$ is not allowed.

\section{The approximate expression of $\hat{\beta}_{B}$ and $\hat{T}_{G}$ in large-$N$ limit} \label{largen}
\setcounter{equation}{0}
Using the Laplace approximation \cite{bib60,bib61} or the Darwin–Fowler method in Sec. 9.1 of \cite{bib62}, $\ln Z$ can be approximated as \footnote{We assume that the system does not undergo a phase transition as the requirement of the Laplace approximation.}
\begin{equation} \label{ln Z}
	\begin{split}
		\ln Z=&\ln \sigma(E^{*})+
		\ln \sqrt{\left. 2\pi\left( -\frac{\partial\hat{\beta}_{B}}{\partial E}\right)^{-1}\right| _{E=E^{*}} }-\beta E^{*}+O\left( N^{-1}\right)  ,
	\end{split}
\end{equation}
where $\left. \partial\ln\sigma(E) / \partial E \right| _{E=E^{*}}=\beta$, i.e. $\hat{\beta}_{B}(E^{*})=\beta$. Here, $E^{*}$ denotes the most probable energy in the canonical ensemble. Then, we obtain the expression average energy $\left\langle E\right\rangle =-\partial \ln Z / \partial \beta$,
\begin{equation} \label{AE}
	\left\langle E\right\rangle =E^{*}+\left. \frac{1}{2} \frac{\partial^{2}\hat{\beta}_{B}}{\partial E^{2}}\right| _{E=E^{*}} \left( \frac{\partial E^{*}}{\partial \beta}\right) ^{2}+O(N^{-1}) .
\end{equation}
Performing the inverse map from $\left\langle  E \right\rangle $ to $E^{*}$ shows the relation between the most probable energy $E^{*}$ and the average energy $\left\langle E\right\rangle $,
\begin{equation} \label{MPE}
	E^{*}=\left\langle E\right\rangle -T+\frac{T^{2}}{2}\frac{\partial \ln C}{\partial T}+O(N^{-1}),
\end{equation}
where $C=\partial \left\langle E\right\rangle / \partial T$ denotes the heat capacity. Substituting Eq. \eqref{MPE} into Eq. \eqref{ln Z}, we have
\begin{equation} \label{entropy}
	\ln \sigma(\left\langle E\right\rangle )=S-\ln \sqrt{ 2\pi\left( -\frac{\partial\left\langle E\right\rangle}{\partial \beta}\right) } +O(N^{-1}),
\end{equation}
where $S=\ln Z-\beta  \partial \ln Z/ \partial \beta $ is the thermodynamics entropy in a canonical ensemble. Eq. \eqref{entropy} shows the nonequivalence between micro-canonical ensemble and canonical ensemble for finite-$N$ systems, i.e., the relation between the Boltzmann entropy $S_{B}=\ln [\sigma(E)\epsilon]$ and the thermodynamic entropy $S(E)$. If we replace $\left\langle E\right\rangle $ with a sample of system's energy $E$, the function $\hat{\beta}_{L}(E)  $ (given by $-\left. \frac{\partial\ln Z}{\partial\beta} \right| _{\beta=\hat{\beta}_{L}(E)}=E $) will be a statistic of $\beta$ \footnote{Note that Eq. \eqref{entropy} is invalid for $E=0$, since the right-hand side of Eq. \eqref{entropy} diverges while the left-hand side of Eq. \eqref{entropy} is a constant. The breakdown of Eq. \eqref{entropy} is consistent with the fact that $\Delta E^{2}=CT^{2}\to 0$ as $E\to 0$.}.   Taking the derivative of Eq. \eqref{entropy} with respect to $E$ on both sides, we have 
\begin{equation} \label{BL}
	\hat{\beta}_{B}(E)=\hat{\beta}_{L}(E)+\frac{1}{2}\frac{\text{d}^{2}_{E}\hat{\beta}_{L}(E)}{\text{d}_{E}\hat{\beta}_{L}(E)}+O(N^{-2}).
\end{equation}
This is the relation between the UMVUE of $\beta$ and the first-order moment $\hat{T}_{L}(E)=\hat{\beta}_{L}(E)^{-1}$ (also the maximum likelihood) estimator of $\beta$.

Performing Taylor expansion of Eq. \eqref{BL} at $\langle E\rangle$ with $\hat{\beta}_{L}\left( \left\langle E\right\rangle \right) =\beta$, and considering $\delta E=E-\langle E\rangle $ as a small quantity, we have
\begin{equation} \label{D2}
	\begin{split}
		\hat{\beta}_{B}-\beta=&\text{d}_{\left\langle E\right\rangle }\hat{\beta}_{L}\left( \left\langle E\right\rangle \right)  \delta E +\text{d}_{\left\langle E\right\rangle }^{2}\hat{\beta}_{L}\left( \left\langle E\right\rangle \right)\frac{(\delta E)^{2}}{2}+\text{d}_{\left\langle E\right\rangle }^{3}\hat{\beta}_{L}\left( \left\langle E\right\rangle \right)\frac{(\delta E)^{3}}{6}\\
		&+
		\frac{1}{2}\frac{\text{d}_{\left\langle E\right\rangle }^{2}\hat{\beta}_{L}\left( \left\langle E\right\rangle \right)}{\text{d}_{\left\langle E\right\rangle }\hat{\beta}_{L}\left( \left\langle E\right\rangle \right)}+\text{d}_{\left\langle E\right\rangle }\left(\frac{\text{d}_{\left\langle E\right\rangle }^{2}\hat{\beta}_{L}\left( \left\langle E\right\rangle \right)}{\text{d}_{\left\langle E\right\rangle }\hat{\beta}_{L}\left( \left\langle E\right\rangle \right)}\right) \frac{\delta E}{2}+O(N^{-2}).
	\end{split}
\end{equation}
Using Eq. \eqref{D2}, we obtain the variance of $\hat{\beta}_{B}$, $\Delta \beta^{2}_{B}=\langle (  \hat{\beta}_{B}-\beta ) ^{2}\rangle $, which can be expanded as
\begin{equation}\label{variance}
	\begin{split}
		\Delta \beta^{2}_{B} 
		=& \left( \text{d}_{\left\langle E\right\rangle}\hat{\beta}_{L}\right) ^{2}\left\langle (\delta E)^{2}\right\rangle + \left( \text{d}_{\left\langle E\right\rangle}\hat{\beta}_{L}\right) \left( \text{d}^{2}_{\left\langle E\right\rangle}\hat{\beta}_{L} \right) \left\langle (\delta E)^{3}\right\rangle \\ &+\left(\text{d}_{\left\langle E\right\rangle}\hat{\beta}_{L}\right)  \text{d}_{\left\langle E\right\rangle}\left( \frac{\text{d}^{2}_{\left\langle E\right\rangle}\hat{\beta}_{L}}{\text{d}_{\left\langle E\right\rangle}\hat{\beta}_{L}}\right) \left\langle  (\delta E)^{2}\right\rangle   + \frac{\text{d}^{2}_{\left\langle E\right\rangle}\hat{\beta}_{L}}{2} \frac{\text{d}^{2}_{\left\langle E\right\rangle}\hat{\beta}_{L}}{\text{d}_{\left\langle E\right\rangle}\hat{\beta}_{L}}\left\langle (\delta E)^{2}\right\rangle \\
		& +\frac{1}{4} \left( \frac{\text{d}^{2}_{\left\langle E\right\rangle}\hat{\beta}_{L}}{\text{d}_{\left\langle E\right\rangle}\hat{\beta}_{L}}\right) ^{2}+ \frac{\left( \text{d}_{\left\langle E\right\rangle}\hat{\beta}_{L}\right) \left( \text{d}^{3}_{\left\langle E\right\rangle}\hat{\beta}_{L}\right) }{3}\left\langle (\delta E)^{4}\right\rangle +\frac{\left( \text{d}^{2}_{\left\langle E\right\rangle}\hat{\beta}_{L}\right) ^{2}}{4}\left\langle (\delta E)^{4}\right\rangle +O(N^{-3}).
	\end{split}
\end{equation}
In the derivation, we have used $\delta E \sim  N^{1/2}$ and $\text{d}_{\left\langle E\right\rangle }\sim N^{-1}$.
To simplify Eq. \eqref{variance}, we introduce the higher order derivatives of the inverse function \cite{bib66}, 
\begin{equation} \label{Taylor}
	\begin{split}
		\text{d}_{\left\langle E\right\rangle}\hat{\beta}_{L}&=\left( \frac{\partial \left\langle E\right\rangle }{\partial \beta}\right) ^{-1}  \\
		\text{d}^{2}_{\left\langle E\right\rangle}\hat{\beta}_{L}&=-\frac{\partial^{2}\left\langle E\right\rangle }{\partial \beta^{2}}\left( \frac{\partial \left\langle E\right\rangle }{\partial\beta}\right) ^{-3}  \\
		\text{d}^{3}_{\left\langle E\right\rangle}\hat{\beta}_{L}&=\left( \frac{\partial \left\langle E\right\rangle }{\partial\beta}\right) ^{-5}\left[ 3\left( \frac{\partial^{2}\left\langle E\right\rangle }{\partial \beta^{2}}\right) ^{2}-\frac{\partial\left\langle E\right\rangle }{\partial \beta}\frac{\partial^{3}\left\langle E\right\rangle }{\partial \beta^{3}}\right].
	\end{split}
\end{equation}
For convenience of calculation, we list the following expressions: $\partial \left\langle E\right\rangle / \partial\beta=-\left\langle (\delta E)^{2}\right\rangle$, $\partial^{2}\left\langle E\right\rangle / \partial \beta^{2}=\left\langle (\delta E)^{3}\right\rangle$ and $\partial^{3}\left\langle E\right\rangle / \partial \beta^{3}=3(\left\langle (\delta E)^{2}\right\rangle)^{2}-\left\langle (\delta E)^{4}\right\rangle$.
Then, substituting Eq. \eqref{Taylor} into Eq. \eqref{variance}, we obtain the variance of $\hat{\beta}_{B}$,
\begin{equation}
	\Delta \hat{\beta}_{B}^{2} =\frac{1}{\left\langle (\delta E)^{2}\right\rangle }+\frac{1}{2}\frac{\left\langle (\delta E)^{3}\right\rangle ^{2}}{(\left\langle (\delta E)^{2}\right\rangle)^{4}}+O(N^{-3}).
\end{equation}

Next, we consider the approximate expression of $S_{G}=\ln \Omega$. Using the relation $\partial \Omega(E)/ \partial E=\sigma(E)$, we have,
\begin{equation} \label{GB}
	S_{G}(E)= S_{B}(E)+\ln \hat{T}_{G}(E)+O(N^{-1}),
\end{equation}
Eq. \eqref{GB} is consistent with the exact relation $\hat{T}_{G}=\hat{T}_{B}(1-C_{G}^{-1})$ , where $C_{G}=\partial E/ \partial \hat{T}_{G}$ \cite{bib1}. Taking the derivative of Eq. \eqref{GB} with respect to $E$ on both sides and using Eq. \eqref{BL}, we have
\begin{equation} \label{GL}
	\hat{T}_{G}(E)=\hat{T}_{L}(E)-\frac{\hat{T}_{L}(E)^{2}}{2}\frac{\text{d}^{2}_{E}\hat{T}_{L}(E)}{\text{d}_{E}\hat{T}_{L}(E)}+O(N^{-2}),
\end{equation}
Following the similar procedure, we obtain the variance of $\hat{T}_{G}$ using Eqs. \eqref{Taylor} \eqref{GL}  and $\hat{T}_{L}\left( \left\langle E\right\rangle \right) =T$,
\begin{equation}
	\Delta T^{2}_{G} =\frac{T^{4}}{\left\langle (\delta E)^{2}\right\rangle}+\frac{T^{12}(\text{d}^{2}_{T}\left\langle E\right\rangle)^{2} }{2(\left\langle (\delta E)^{2}\right\rangle)^{4}}+O(N^{-3}).
\end{equation}

\section{Estimation of temperature for the classical ideal gas} \label{classical}
\setcounter{equation}{0}
In the classical ideal gas, the density of states is $\sigma(E)=E^{3N/2-1}$ \cite{bib62}. We obtain the partition function of the system 
\begin{equation}\label{appendix par}
	Z=\int_{0}^{\infty}\sigma(E)\text{e}^{-\beta E}=\beta^{-\frac{3N}{2}}\Gamma\left( \frac{3N}{2}\right) ,
\end{equation}
the most probable energy
\begin{equation}
	E^{*}=\frac{3N-2}{2\beta},
\end{equation}
the Boltzmann temperature and the Gibbs temperature
\begin{equation}
	\hat{\beta}_{B}=\frac{3N-2}{2E},
	\hat{T}_{G}=\frac{2E}{3N} .
\end{equation}

Using Eq. \eqref{appendix par}, we obtain the average energy
\begin{equation} \label{appen aver}
	\left\langle E\right\rangle =-\frac{\partial\ln Z}{\partial\beta}=\frac{3N}{2\beta},
\end{equation}
the heat capacity
\begin{equation} \label{appen cap}
	C=\frac{\partial \left\langle E\right\rangle}{\partial T}  =\frac{3N}{2},
\end{equation}
and the first moment estimator of the temperature (inverse temperature)
\begin{equation}
	\hat{T}_{L}(E)=\frac{2E}{3N},\quad \hat{\beta}_{L}(E)=\frac{3N}{2E}.
\end{equation}
Substitute the above expressions into both sides of Eqs. \eqref{entropy12}, \eqref{BL3} and \eqref{GL3}. We have
\begin{equation} \label{rhs}
	\begin{split}
		\left( \frac{3N}{2}-1\right) \left( \ln\frac{3N}{2}-\ln\beta\right) 
		=&\left(1-\frac{3N}{2}\right) \ln \beta +\ln\Gamma\left( \frac{3N}{2}\right) -\frac{1}{2}\ln2\pi\\
		&-\frac{1}{2}\ln\left( \frac{3N}{2}\right)+\left( \frac{3N}{2}\right)+ O(N^{-1}),
	\end{split}	
\end{equation}
where we have used the String approximation \cite{arfken2011mathematical} in the derivation.
Moreover, the left-hand side (LHS) and the right-hand side (RHS) of Eqs. \eqref{BL3} and \eqref{GL3} are exactly equal, because our model is sufficiently simple. 

The probability distribution of the system’s energy is a gamma distribution 
\begin{equation} \label{appen gamma}
	P(E)=\frac{\beta^{\frac{3N}{2}}E^{\frac{3N}{2}-1}\text{e}^{-\beta E}}{\Gamma\left( \frac{3N}{2}\right) }.
\end{equation}
Using Eq. \eqref{appen gamma}, we obtain the standard deviation of Boltzmann temperature
\begin{equation} \label{appen stan bolt}
	\Delta \hat{\beta}_{B}=\beta\sqrt{\frac{2}{3N-4}},		
\end{equation}
the standard deviation of energy
\begin{equation} \label{appen stan energy}
	\Delta E=\beta^{-1}\sqrt{\frac{3N}{2}},
\end{equation}
and the skwness of this disribution
\begin{equation}\label{appen kapabeta}
	\mu_{3}=\sqrt{\frac{8}{3N}}.
\end{equation}

Thus, the result of the achievable ETU (Eq.~\eqref{Dbetayange}) using Eqs. \eqref{appen stan bolt} \eqref{appen stan energy} is:
\begin{equation} \label{AETUZM}
	\begin{split}
		\Delta \hat{\beta} \Delta E  & \geq \Delta \hat{\beta}_{B} \Delta E =\sqrt{\frac{3N}{3N-4}}>1.
	\end{split}
\end{equation}
And the refined ETU (Eq.~\eqref{Dbeta}) is expressed as
\begin{equation} \label{RETUZM}
	\begin{split}
		\sqrt{\frac{3N}{3N-4}}  &=1+\frac{2}{3N}+O(N^{-2}).
	\end{split}
\end{equation}
Following the similar procedure, we obtain
\begin{equation}
	\Delta \hat{T}_{G}=\beta^{-1}\sqrt{\frac{2}{3N}},\quad
	\Delta E=\beta^{-1}\sqrt{\frac{3N}{2}},
\end{equation}
and 
\begin{equation}
	C=\frac{\partial \left\langle E\right\rangle}{\partial T}  =\frac{3N}{2}.
\end{equation}
Then, Eqs.~\eqref{DT33} and~\eqref{DT} are given straightforwardly.

\end{document}